\documentclass[prb,aps,amsmath,12pt,a4paper,showpacs]{revtex4}

\usepackage{amsfonts,epsfig}

\begin{document}

\title{Supersymmetry Approach to Almost Diagonal Random Matrices  }

\author{Oleg Yevtushenko}

\affiliation{The Abdus Salam ICTP,  Strada Costiera 11,
                   I-34014 Trieste, Italy}

\email{bom@ictp.it}

\author{Alexander Ossipov}

\affiliation{Instituut-Lorentz, Universiteit Leiden, P.O. Box 9506,
                  2300 RA Leiden, The Netherlands}

\email{ossipov@lorentz.leidenuniv.nl}

\begin{abstract}

We  develop a supersymmetric field theoretical description of
the Gaussian ensemble of the almost diagonal Hermitian Random 
Matrices. The matrices have independent random entries $ \,
H_{i \ge j} \, $ with parametrically small off-diagonal elements $ \,
{H_{ij}/H_{ii} \sim {\cal B} \ll  1} $.  We derive a regular virial 
expansion of correlation functions in the number of ``interacting'' 
supermatrices associated with different  sites in the real space 
and demonstrate that the perturbation theory constructed in this 
way is controlled by a small parameter ${\cal B}$. General form of the 
integral expression for the $ \, m $-th virial coefficient governed 
by the ``interaction'' of $ \, m \, $ supermatrices is presented 
and calculated explicitly in the cases of 2- and 3-matrix ``interaction''. 
The suggested technique allows us to calculate both the spectral 
correlations and the correlations of the eigenfunctions taken at 
different energies and in different space points.
\end{abstract}

\date{\today}

{}


\maketitle

\section{Introduction}
\subsection{Conventional and unconventional Random Matrix Theories}

The Random Matrix Theory (RMT) is a very useful mathematical formalism which 
allows to describe universal properties of complex quantum systems. Let us 
consider an ensemble of $ \, N \times N \, $ Hermitian matrices, whose
elements are independent Gaussian distributed random  variables with a zero
mean value and a position dependent variance: 
\begin{equation}\label{RMTdef}
   \langle H_{ij} \rangle = 0 \, ; \quad
   \langle H_{ii}^2 \rangle = \frac{1}{\beta} \, , \quad
   \langle | H_{i \ne j} |^2 \rangle = \frac{1}{2} \, {\cal B}^2 \, {\cal F}( | i - j | ) \, .
\end{equation}
Here $ \, \langle \ldots \rangle \, $ denotes averaging over different realizations of 
RMs; the parameter $ \, \beta \, $ corresponds to the Wigner--Dyson symmetry 
classes: $ \, \beta = 1 \, $ for the Gaussian orthogonal ensemble (real matrices) 
and $ \, \beta = 2 \, $ for the Gaussian unitary ensemble (complex matrices). The 
function $ \, {\cal F} \, $ and the parameter $ \, {\cal B} \, $ determine 
various universality classes sharing the same global symmetries.

A special case of the constant variance of the off-diagonal elements $ \, 
{\cal B}^2 \, {\cal F}( | i - j| ) = 1$ corresponds to the archetypal Wigner--Dyson 
RMT [\onlinecite{Mehta}]. It has a great number of applications starting from nuclear 
physics to quantum chaos to mesoscopic physics \cite{RMT-Rev,Dots-Rev}. 
Ergodic wavefunctions and a level repulsion are essential features 
of the Wigner--Dyson RMT.

Recently, {\it unconventional} RMTs characterized by decreasing 
function $ \, {\cal F} \, $ have attracted a substantial interest (see, for instance, 
detailed introductions in references [\onlinecite{Trotter,DOS,Chi-Virial}]). 
This interest is stimulated, in  particular, by a possibility to explore 
the properties of localized and critical disordered systems. For example, 
if the off-diagonal matrix elements are essentially non-zero only inside 
a band centered at the main diagonal and decay exponentially fast to zero 
outside the band, all eigenfunctions are exponentially localized. This {\it 
banded} RMT describes the physics of a quasi-one-dimensional disordered 
wire \cite{MF-BRM}. If $ \, {\cal F} \, $ decays only as a power-law 
 outside the band \cite{MF-PLBRM}, $ \, {\cal F} \sim 1/| i - j |^{2  \alpha} $, 
the eigenfunctions are power-law localized for $\, \alpha > 1 $. The
Wigner--Dyson universality class is  approached for $ \, \alpha \to 0 $. 
Thus, the {\it power-law banded} RMT (PLBRM) can interpolate between the
Wigner--Dyson statistics and the Poisson statistics of localized 
system. The special case $ \, \alpha = 1 \, $ corresponds to a critical behavior
similar to that found at the point of the Anderson metal-insulator transition 
 \cite{MF-PLBRM,EvMirl,KrTs}. The function $ \, {\cal F}_{crit} \, $
for the one-parametric family of the critical PLBRM can be defined as follows:
$ \, {\cal F}_{crit} ( | i - j | ) = 1 / \left( {\cal B}^2 + (i-j)^2 \right) $ .
The eigenfunctions of this critical model remain multifractal 
at any $ \, {\cal B} \, $ ranging from the weak multifractality at the 
large band-width $ \, {\cal B} \gg 1$, to the strong 
multifractality for the {\it almost diagonal} RMT $ \, {\cal B} \ll 1$. 

\subsection{From large to small bandwidth RMT: $\sigma$-model versus virial expansion}

The considerable progress in the banded and the power-law banded RMT 
has become possible due to the mapping \cite{MF-BRM,MF-PLBRM} onto 
the nonlinear supersymmetric $ \, \sigma $-model \cite{Efetov}, which is  
a powerful filed-theoretical description of various averaged correlation 
functions. However, such a mapping is only justified in the large bandwidth 
limit. This limitation comes  from the saddle-point approximation which is 
a crucial step in the derivation of the $\sigma$-model. Physically
it corresponds to the diffusive approximation which implies in particular 
that the smooth envelope of a typical eigenfunction changes slowly on the
scale of the mean free path. This approximation fails in the opposite limit 
of the small bandwidth, including the case of the {\it almost diagonal RMTs} 
where the bandwidth shrinks to zero and the off-diagonal matrix elements are 
assumed to be parametrically smaller than the diagonal ones, $ \, {\cal B} \ll 1 $. 

Let us represent the almost diagonal matrix as a sum of the diagonal part and a matrix 
of the small off-diagonal elements $ \, \hat{H} = \hat{H}_d + \hat{V} \, , \  V \sim {\cal B} 
\ll 1 $. The diagonal matrix $ \, \hat{H}_d \, $ represents ``non-interacting'' energy 
levels or localized eigenstates. The presence of the small off-diagonal matrix 
$ \, \hat{V} \, $ leads to a weak ``interaction'' between different localized states. 
In order to calculate correlation functions, one can perform an expansion  in 
the number of interacting localized states \cite{Levitov}. The small parameter $ \, 
{\cal B} \, $ is the control parameter of this procedure. This method 
was called \cite{Trotter,DOS,Chi-Virial} \ ``{\it a virial expansion}'' (VE) by analogy 
with the expansion of thermodynamic functions of a dilute system in powers of 
density with the $ \, m $-th virial coefficient being governed by collisions of $ \, 
m + 1 \, $ particles.

These ideas were initially implemented in  a semi-empirical real-space 
renormalization group approach which has been applied for critical systems 
with long-range interactions \cite{Levitov} and for a quantum Kepler problem 
\cite{AltLev}. The real-space renormalization group was also used to study the 
critical almost diagonal PLBRM \cite{EvMirl}: the spectral correlations and  the
scaling properties of the eigenfunctions were investigated by considering a resonant 
interaction of two energy levels. The renormalization group approach has however 
two serious disadvantages: (i)~it does not allow for a rigorous control of omitted 
contributions; (ii)~due to technical difficulties, it is almost impossible to go beyond 
the leading term, thus  the resonant interaction of three  and more levels remains 
inaccessible in this framework. 

A rigorous counterpart of the renormalization group approach has been
suggested in recent works [\onlinecite{Trotter,DOS,Chi-Virial}]. It deals
with a {\it regular} VE generated with the help of the Trotter formula (TVE) 
\cite{Trt}. Based on the classification of perturbation series by a number of 
the interacting levels involved, TVE allows one to study the density of states 
and spectral correlations of the almost diagonal RMTs.
The accuracy of the TVE is always controllable resulting in the rigorous 
perturbation theory. The second disadvantage of the renormalization group 
has been also partly overcome: the TVE allows to go beyond the leading term by 
considering the interaction of two and three levels. In this way the first and 
the second virial coefficients have been calculated for a generic model of the 
almost diagonal RMTs. The general formulas have been applied 
\cite{Trotter,DOS,Chi-Virial} to the different models of almost diagonal 
RMTs, including critical PLBRMs \cite{MF-PLBRM}, the unitary 
Moshe-Neuberger-Shapiro model \cite{MNS}, and the Rosenzweig-Porter 
model \cite{RosPort} in the regime of crossover.

The TVE involves a complicated combinatorial part of intermediate calculations. 
The combinatorial problem appeared there resembles a coloring problem 
of closed graph edges. If one considers the density of states \cite{DOS}, 
the coloring is similar to the well studied problem of the graph theory
\cite{Graphs} whose solution is known \cite{BK}. However a study of 
spectral correlations requires to resolve much more complicated problem
of simultaneous coloring of several graphs \cite{Trotter}. A complexity of 
the combinatorial calculations grows tremendously with increasing the 
number of the interacting energy levels. Therefore the TVE can be used  
practically only for calculation of the first and the second virial coefficients. 

\subsection{Virial expansion from the field theoretical representation}

In the present work we formulate a {\it supersymmetric field-theoretical 
representation for the VE} of different correlation functions of the 
generic model (\ref{RMTdef}). The method of the supersymmetry
allows us to perform an averaging over RM ensemble for an arbitrary
function $ \, {\cal F}( | i - j | ) $.  The supersymmetric VE is controlled 
by a small parameter $ \, {\cal B} \ll 1 \,$. The virial coefficients are straightforwardly
derived in a general form in terms of the integrals over supermatrices.
It is important that no combinatorics appear in the intermediate field-theoretical 
calculations. In this framework, the interaction of $ \, m \, $ energy levels is 
described by an integral containing only  $ \, m \, $ independent
supermatrices associated with $ \, m \, $ different sites in the real space. 
In order to calculate the integrals over supermatrices explicitly we employ a 
parametrization introduced recently in Ref.[\onlinecite{SashaKr}].

The suggested  supersymmetric field theory (SuSyFT) can be 
equally applied  both to the spectral correlations and to  the correlations of 
the eigenfunctions taken at different energies and in different space points. 
We emphasize that  this approach  is the unique analytic tool to describe the 
wavefunctions correlations for the almost diagonal RMTs. Neither the real space 
renormalization group nor the TVE are capable
to do this. We would like to mention also that SuSyFT might be useful
for non-perturbative calculations as well, however this issue is beyond
the scope of the present work. 

The paper is organized as follows: we present the SuSyFT
and give main definitions in the Section \ref{SectMainDef}. The basic
ideas and parameters of the VE are explained in the Section \ref{SectVE}
with a reference to SuSyFT. A general integral expression for 
$ \, m $-th virial coefficient governed by the interaction of $ \, m \, $
supermatrices  is presented in the Section \ref{SectVirCoef}.
In this section we also discuss in detail the validity of a saddle-point 
integration over massive degrees of freedom and the applicability
of the VE. We exemplify the integral calculations for the cases of 2- and 3-matrix 
interaction in the Section \ref{Sect2and3Q} and discuss advantages of the
method and possible further  applications in Conclusions.

\section{Main definitions}\label{SectMainDef}

Let us introduce the retarded and advanced Green's functions
\begin{equation}\label{Greens}
   \hat{G}^{R/A}(E) = \frac{1}{E - \hat{H} \pm \imath 0} \, ;
\end{equation}
$ \, \hat{H} \, $ is a Hermitian RM of large size $ \, N \gg 1 \, $.
It has independent matrix elements and belongs to the Gaussian 
ensemble described by Eq.(\ref{RMTdef}). Without loss of generality,
we  consider the case of GUE, $ \, \beta = 2 $. A generalization
to  other symmetry classes is straightforward. The Green's functions
are an efficient tool to study different correlation functions. For 
example, the expression for the averaged density of states in terms 
of the Green's functions reads
\begin{equation}\label{rho}
    \langle \rho(E) \rangle = \frac{1}{N}\sum_{n=1}^{N} \langle 
                                                     \delta( E - \epsilon_n ) \rangle =
    \frac{1}{\pi N} \Im \, \left\langle 
                     {\rm Tr} \left( \hat{G}^{A}(E) \right) 
                                     \right\rangle,
\end{equation}
where $ \, \epsilon_n \, $ are eigenvalues of the random matrix $ \, \hat{H} 
\, $ and $ \, \langle \ldots \rangle \, $ denotes averaging over the ensemble 
of random matrices. The inverse density of states taken at the band center $ \, E = 0 \, $
governs the mean level spacing of RMT: $ \, \Delta = 1/N\langle \rho(0) \rangle
$. The mean level spacing of almost diagonal unitary RMT is \cite{DOS}
\[
   \Delta \Bigl|_{{\cal B} \ll 1} \simeq  ( \sqrt{\pi} + O({\cal B}^2) ) N^{-1} \, .
\]
The two-point correlation functions can be expressed by 
means of the quantity $ \, {\cal G}_{pq} ( \omega ) $:
\begin{equation}\label{Gpq}
    {\cal G}_{pq}( \omega ) \equiv \frac{1}{\Delta} \int_{-\infty}^{\infty} \!\!\! {\rm d } E \
         \hat{G}^R_{pp} (E + \omega/2 ) \hat{G}^A_{qq} (E - \omega/2 ) \, .
\end{equation}
For example, the expressions for the  averaged two-level correlation function
\begin{equation}\label{R2}
    R_2(\omega) \equiv {N\Delta} \int_{-\infty}^{\infty} \!\!\! {\rm d } E
                   \langle \langle 
                            \, \rho( E + \omega/2) \rho( E - \omega/2) \, 
                   \rangle \rangle 
\end{equation}
is given by 
\begin{equation}\label{R2-int}
    R_2(\omega) = 
       \frac{\Delta^2}{2 \pi^2 N} \, \Re \sum_{p,q=1}^{N}
                   \langle \langle \, 
                           {\cal G}_{pq}( \omega )
                   \, \rangle \rangle, 
\end{equation}
where $ \, \langle \langle a b \rangle \rangle \equiv \langle a b \rangle
- \langle a \rangle \langle b \rangle $; while the averaged
correlator of two eigenfunctions taken at different energies at
different space points $ \, p \mbox{ and } q \, $
\begin{equation}\label{C2}
    C_2(\omega,p,q) \equiv \Delta \int_{-\infty}^{\infty} \!\!\! {\rm d } E \, 
                   \langle \langle \,
                       \sum_{m,n=1}^N 
                       \delta( E +\omega/2 - \epsilon_n ) \delta( E -\omega/2 - \epsilon_m )
                      | \psi_{\epsilon_n}(p) |^2 | \psi_{\epsilon_m}(q) |^2
                   \, \rangle \rangle
\end{equation}
reads
\begin{equation}\label{C2-int}
    C_2(\omega,p,q) = 
      \frac{\Delta^2}{2 \pi^2 N} \, \Re \, 
                   \langle \langle \,
                       {\cal G}_{pq}( \omega ) 
                   \, \rangle \rangle .
\end{equation}
Thus, we have to calculate $ \, \langle \langle \, {\cal G}_{pq} (\omega ) \, \rangle 
\rangle \, $ to explore the two-point correlation functions. The ensemble averaging can be 
performed with the help of SuSyFT [\onlinecite{Efetov,MirlinLect}]. To this end, we introduce 
$ \, N \, $  supervectors
\begin{eqnarray}\label{S-vevtors}
   \Psi(\alpha) = \left(
                \begin{matrix} 
                    \Psi^R(\alpha) \\ \Psi^A(\alpha)
                \end{matrix}
                             \right) \, ; \quad 
   \alpha = 1, 2, \ldots N \, .
\end{eqnarray}
Here, $ \, \Psi^{R/A}(\alpha) = ( s^{R/A}(\alpha) , \chi^{R/A}(\alpha) )^T $
are supervectors in retarded-advanced sectors. They consist of commuting 
$ \,  s^{R/A}(\alpha) \, $ and anti-commuting Grassmann variables $ \, \chi^{R/A}
(\alpha) $. The direct product of $ \, \Psi(\alpha) $-vectors with the conjugated vectors 
constitutes $ \, N \ Q $-matrices of the size $ \, 4 \times 4 $. The block structure of the 
$ \, Q $-matrices in the retarded-advanced notation is:
\begin{eqnarray}\label{Q-blocks}
    Q_{\alpha} \equiv \Psi (\alpha) \otimes \bar{\Psi}(\alpha) = 
      \left( 
          \begin{matrix} 
        \Psi^R(\alpha) \otimes \left( \Psi^R(\alpha) \right)^\dagger & 
        \Psi^R(\alpha) \otimes K \left( \Psi^A(\alpha) \right)^\dagger 
                       \\ 
        \Psi^A(\alpha) \otimes \left( \Psi^R(\alpha) \right)^\dagger  & 
        \Psi^A(\alpha) \otimes K \left( \Psi^A(\alpha) \right)^\dagger
          \end{matrix}
      \right) \! , \ 
    K \equiv \left( \begin{matrix} 
          -1 & 0 \\
           0 & 1
                            \end{matrix}
                  \right) \! .
\end{eqnarray}
The ensemble averaged $ \,  {\cal G}_{pq} \, $ can be written as follows
\begin{eqnarray}\label{RA-SuSy}
      \left\langle \,
         {\cal G}_{pq} (\omega )
      \, \right\rangle = \frac{(-1)^{N+1}}{\Delta} \int_{-\infty}^{\infty} \!\!\! {\rm d } E \,
   \int {\cal D} \{ Q \} \, 
                 {\cal R}_{p} \, {\cal A}_{q} \,
             \left( \prod_{\alpha = 1}^N e^{S_0[Q_{\alpha}]} \right)
             \left( \prod_{n \ne m}^N e^{S_p[ Q_n , Q_m ]} \right) \, ;
\end{eqnarray}
where $ \, {\cal D} \{ Q \} \equiv \prod_{\alpha = 1}^N {\cal D} \{ Q_{\alpha} \} \, $
is the  measure of integration over the supermatrices $ \, Q_{\alpha}
$ (see Appendix \ref{Q-param} for the details) \cite{Psi-Q}. The factors $ \,  {\cal R}_p \, $ 
and $ \,  {\cal A}_q \, $  break the supersymmetry between the commuting and
anti-commuting variables in the retarded/advanced sectors:
\begin{equation}
   {\cal R}_{p} = \left( \chi^R(p) \right)^* \chi^R(p) \, , \quad
   {\cal A}_{q} = \left( \chi^A(q) \right)^* \chi^A(q) \, .
\end{equation}
In Eq.(\ref{RA-SuSy}) we have separated out two parts of the action: $
\, S_0 \, $  corresponds to the diagonal part of RM and depends on a single supermatrix
\begin{eqnarray}\label{S0-def} 
   S_0[Q_\alpha] & = & {\rm Str} \left\{
                   - a \, Q_\alpha^2 + \imath \left( E + \frac{\Omega}{2} \Lambda \right) Q_\alpha
                                          \right\} \, , \quad 
     \Lambda \equiv \left( \begin{matrix} 
          1 & 0 \\
          0 & -1
                            \end{matrix}
                  \right)_{RA} \! , \\
\label{S0-def-add} 
     a & \equiv & \frac{1}{2} \langle H_{\alpha\alpha}^2 \rangle = \frac{1}{4} \, , \quad
     \Omega \equiv \omega + \imath \, 0 \, . 
\end{eqnarray}
The second part of the action $ \, S_p \, $ is proportional to the variance of the
off-diagonal elements and contains a product of two supermatrices
\begin{eqnarray}\label{Sp-def} 
   S_p[ Q_k , Q_m ] & = & - b_{km} {\rm Str} \left\{ Q_k Q_m \right\} \, , \quad k \ne m \, , \\
\label{Sp-def-add}
       b_{km} & \equiv & \frac{1}{2} \langle |H_{km}|^2 \rangle = 
                                             \frac{1}{4} \, {\cal B}^2 \, {\cal F}( | k - m | ) \, .
\end{eqnarray}
In the other worlds, $ \, S_p \, $ describes an interaction of different $ \, Q $-matrices.
Note that Eqs.(\ref{RA-SuSy},\ref{S0-def}--\ref{Sp-def-add}) are exact. We have used 
$ \, Q $-matrices to compactify the notation but we could equally write the expression for $ \, \left
\langle \, {\cal G}_{pq} (\omega ) \, \right\rangle \, $ in terms of the integrals over $ \, \Psi $-vectors.

The standard $ \, \sigma $-model derivation includes (i)  introducing
auxiliary supermatrices $ Q_{\alpha}^{(\sigma)}$  that allow to decouple the quartic 
in $ \, \Psi_{\alpha} \, $  (i.e., quadratic in $ \, Q_{\alpha} $) terms by the Hubbard-Stratonovich 
transformation; (ii) Gaussian integration over the supervectors $ \, \Psi_{\alpha} $. After these 
two steps,  one has to employ the saddle-point approximation for the integrals over $ \, 
Q_{\alpha}^{(\sigma)} $.  As we have already discussed in the Introduction, this procedure is
justified  only for the large band width RMT, $ \, {\cal B} \gg 1 $. Here in
contrast we will consider the case
\begin{equation}\label{SmallB}
    {\cal B} \ll 1 \, ,
\end{equation}
where the standard $ \, \sigma $-model fails. Therefore instead of the Hubbard-Stratonovich 
transformation with further standard steps, we will use the method of the virial expansion
and employ a parametrization of the $ \, Q $-matrices suggested in the reference
[\onlinecite{SashaKr}]. Details of the parametrization are given in Appendix \ref{Q-param}.
We emphasize that integration manifolds of  the $ \, \sigma $-model
and of the VE {\it are quite different}: it is given by $\frac{U(1,1)}{U(1)\times
U(1)}\times \frac{U(2)}{U(1)\times U(1)}$  for the former and by  $\frac{U(1,1)}{U(1)\times
U(1)}\times  \mathbb{R}$ for the latter \cite{SashaKr}. In particular, a saddle-point
approximation in the SuSyFT suggested will lead to the linear constraint
$ \, {\rm Str}(Q) = 0 \, $ for a large scale theory (see Sect. \ref{SectVirCoef}) whereas 
the saddle-point manifold of the standard diffusive $ \, \sigma $-model requires the 
additional nonlinear constraint $ \, \bigl( Q^{(\sigma)} \bigr)^2 = 1 $. Notice however
that the nonlinearity of SuSyFT follows already from the definition of the matrix
$Q\equiv \Psi \otimes \bar{\Psi}$.

\section{Basic concept of the virial expansion}\label{SectVE}

\begin{figure}[t]
\unitlength1cm
\begin{picture}(13.5,11)
   \epsfig{file=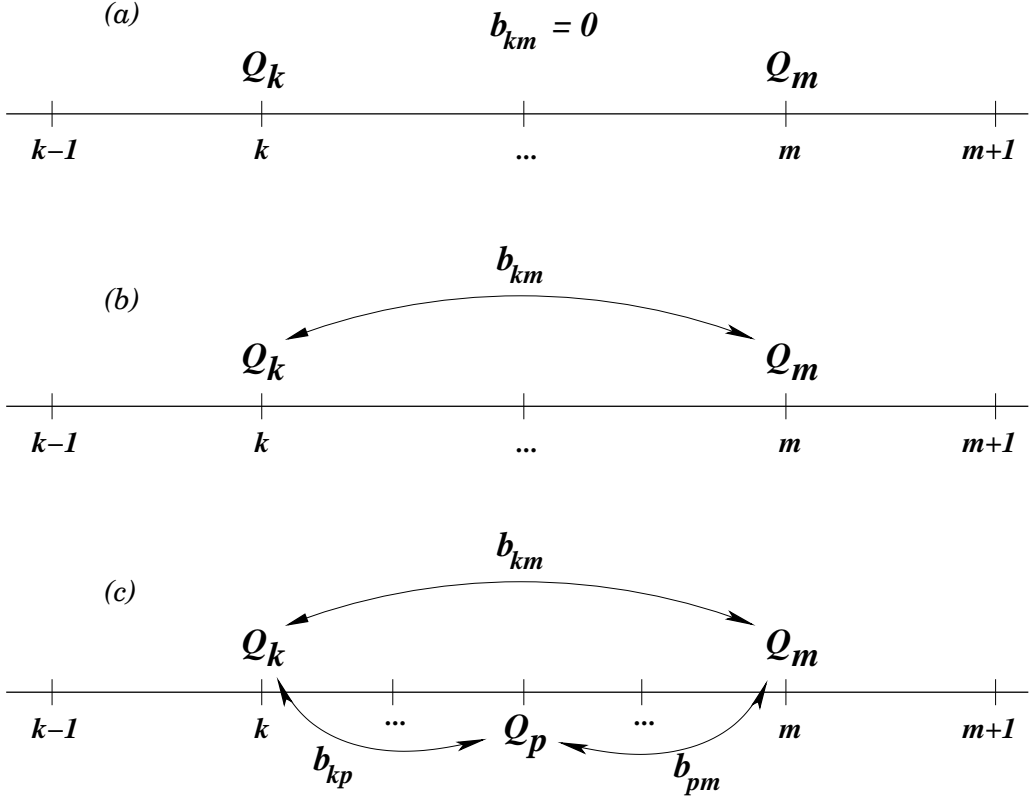,angle=0,width=13.5cm}
\end{picture}
\vspace{0.5cm}
\caption{
\label{VE-picture}
Scheme of the SuSyFT virial expansion: (a) Supermatrices with different indices
(associated with different sites in the real space) do not interact with each 
other in the case of the diagonal RMT possessing the localized wavefunctions; 
(b) Pair interaction of of the supermatrices via (small) off-diagonal elements
of RMs, which governs the first virial coefficient $ \, \bar{{\cal G}}^{(2)} $, see
Eqs.(\ref{VE-coef-def},\ref{FuncSer}); (c) The interaction of three supermatrices, 
which governs the second virial coefficient $ \, \bar{{\cal G}}^{(3)} $, see
Eqs.(\ref{VE-coef-def},\ref{FuncSer}).
}
\end{figure}

The two-fold product of exponentials in Eq.(\ref{RA-SuSy}) can be expanded
in power series
\begin{equation}\label{ProdSum}
   \left( \prod_{n \ne m}^N e^{S_p[ Q_n , Q_m ]} \right) =
        \sum_{k=0}^{\infty} \frac{1}{k!}
                        \left(
                           \sum_{m > n=1 }^{N}
                                \Bigl( - 2 b_{m,n} {\rm Str} [Q_m Q_n] \Bigr)
                        \right)^k
     \equiv  {\cal V}^D + 
     \sum_{m=2}^{\infty} {\cal V}^{(m)} \, ; \quad
     {\cal V}^D = 1 \, .
\end{equation}
We have rearranged this series by separating out terms $ \, {\cal V}^{(m)} \, $
which contain a given number $ \, m \ge 2 \, $ of different $ \, Q $-matrices.
It is easy to show that
\begin{eqnarray}\label{Vir2}
   {\cal V}^{( 2)} & = & \sum_{\alpha_1 > \alpha_2 =1}^{N} {\cal V}^{( 2)}_{\alpha_1 \alpha_2} \, , \quad
   {\cal V}^{( 2)}_{\alpha_1 \alpha_2} \equiv
                                        e^{ - 2 b_{\alpha_1 \alpha_2} {\rm Str} [Q_{\alpha_1} Q_{\alpha_2}] } - 1 \, ; \\
\label{Vir3}
   {\cal V}^{( 3)} & = & \!\!\!\!\! \sum_{\alpha_1 > \alpha_2 > \alpha_3 =1}^{N} 
                                {\cal V}^{( 3)}_{\alpha_1 \alpha_2 \alpha_3} \, , \\ 
      & &
   {\cal V}^{( 3)}_{\alpha_1 \alpha_2 \alpha_3} \equiv 
        {\cal V}^{( 2)}_{\alpha_1 \alpha_2} {\cal V}^{( 2)}_{\alpha_1 \alpha_3} {\cal V}^{( 2)}_{\alpha_2 \alpha_3} +
                                           {\cal V}^{( 2)}_{\alpha_1 \alpha_2} {\cal V}^{( 2)}_{\alpha_1 \alpha_3} +
                                           {\cal V}^{( 2)}_{\alpha_1 \alpha_2} {\cal V}^{( 2)}_{\alpha_2 \alpha_3} +
                                           {\cal V}^{( 2)}_{\alpha_1 \alpha_3} {\cal V}^{( 2)}_{\alpha_2 \alpha_3} \, ;
\nonumber
\end{eqnarray}
and so on for the higher terms. The function $ \, {\cal V}^{( 2)}_{\alpha_1 \alpha_2} \, $ in VE is
a counterpart of the Mayer's function used in the theory of imperfect gases \cite{StatPhys}.

This expansion can be now substituted into the expression for 
 $ \, \left\langle \, {\cal G}_{pq} (\omega ) \, \right\rangle \, $:
\begin{equation}\label{RA-VE}
      \left\langle \,
         {\cal G}_{pq} (\omega )
      \, \right\rangle = \frac{(-1)^{N+1}}{\Delta} \int_{-\infty}^{\infty} \!\!\! {\rm d } E \,
   \int {\cal D} \{ Q \} \,
                 {\cal R}_{p} \, {\cal A}_{q} \,
             \left( \prod_{\alpha = 1}^N e^{S_0[Q_{\alpha}]} \right)
             \left(  {\cal V}^D + {\cal V}^{( 2)} + {\cal V}^{(3)} + \ldots \right) .
\end{equation}

The first term $ \, {\cal V}^D = 1 \, $ corresponds to the diagonal RMT with 
noninteracting localized eigenstates. This is the starting point for VE. The wavefunctions 
of the diagonal RMs are completely localized at different sites having no overlap with the 
other sites and the energy levels are uncorrelated:
\begin{eqnarray}
\label{DiagCorr-1}
    \langle\langle \, {\cal G}_{p \ne q}^D \, \rangle\rangle & = & 0 \, , \\
\label{DiagCorr-2}
    \langle\langle \, {\cal G}_{pp}^D \, \rangle\rangle & = & 
        \displaystyle \frac{\pi}{\Delta} \left( \frac{2 \, \imath}{\Omega} -
            e^{- \frac{\omega^2}{2}}
            \int {\rm d} E \, e^{ - 2 E^2 } 
                             \left( {\rm erfi} \left[ E + \frac{\omega}{2} \right] - \imath \right)
                             \left( {\rm erfi} \left[ E - \frac{\omega}{2} \right] + \imath \right) 
                                                              \right) \, ; \\
\label{DiagCorr-3}
    R_2^D |_{N\to\infty} & = &
    \displaystyle  
        \Re \Bigl( \frac{\Delta^2}{2 \pi^2} 
                      \langle\langle \, {\cal G}_{p p}^D (\omega) \, \rangle\rangle 
               \Bigr) = 
                   \Delta \Bigl( \delta(\omega) + O(1) \Bigr) = 
                   \delta(s) + O(1/N) \, ; \\ 
    s & \equiv & \frac{\omega}{\Delta} \, , \quad
   {\rm erfi} (z) \equiv \frac{2}{\sqrt{\pi}} \int_0^z e^{t^2} {\rm d} t \, .
\nonumber
\end{eqnarray}
This is reflected by the structure of the first term: the $ \, Q $-matrices are decoupled 
and all superintegrals factorize. We can say that the matrix $ \, Q_{\alpha} \, $ is associated 
with  the site $ \, \alpha \, $ and the supermatrices at different sites do not 
interact with each other if we put $  \, S_p = 0 $ (see Fig.\ref{VE-picture}a).

Each term $ \, {\cal V}^{( m)} \, $ is the sum of the exponentials containing $ \, m \, $ 
supermatrices.  The supermatrices  are linked by the small off-diagonal elements of RMs 
(see Fig.\ref{VE-picture}b,c). We refer to these links as ``the interaction of the supermatrices''. 
Obviously, they reflect the interaction between the localized eigenstates of the diagonal 
part of RMT. 

Let us discuss general properties of the summands $ \, {\cal V}^{( m)}_{\{ \alpha \}}, \
\{ \alpha \} \equiv \alpha_1, \alpha_2, \ldots , \alpha_m $. The set $ \, \{ \alpha \} \, $
must include external indices ($ \, p \ne q \, $ and $ \, p = q \, $ in the off-diagonal case
and in the diagonal one, respectively); otherwise the contribution $ \, {\cal V}^{( 
m)}_{\{ \alpha \}} \, $ is canceled by subtracting the decoupled term, see the definition 
of $ \, \langle \langle \ldots \rangle \rangle $. The $ \, N - m \, $ supermatrices, whose 
indices do not belong to the set $ \, \{ \alpha \} $, are included neither in $ \, {\cal V}^{(m)} 
\, $ nor in the symmetry breaking factor $ \, {\cal R}_p {\cal A}_q \, $ and, thus, are containing 
only in "non-interacting"  part of the action $ \, S_0 $. The integrals over these supermatrices  
are equal to unity due to the supersymmetry.

Let us introduce several definitions. We will call supermatrices $ \, Q_{\beta_1} \, $ 
and $ \, Q_{\beta_2} \, $ entering into the expression for $ \, {\cal V}^{( m)}_{\{ \alpha 
\}} \, $ {\it connected}, if and only if there is a sequence of indices $\{\gamma_1,\:\dots,\:,
\gamma_n\}$  such that all elements  $b_{\beta_1\gamma_1},\:b_{\gamma_1
\gamma_2},\:\dots,\:b_{\gamma_{n-1} \gamma_n},\:b_{\gamma_{n}\beta_2}$ are 
contained in $ \, {\cal V}^{( m)}_{\{ \alpha \}} \, $. We  refer to (i) a subset $ \, 
\{ \alpha' \} \subset \{ \alpha \} \, $ as to {\it connected subset} if all independent matrices 
$ \, Q_{ \{ \alpha' \} } \, $ are connected to each other; (ii) two non-intersecting 
subsets $ \, \{ \alpha^{(1)} \} \subset \{ \alpha \} $, $ \, \{ \alpha^{(2)}\} \subset 
\{ \alpha \} $, $ \, \{ \alpha^{(1)} \} \bigcap \{ \alpha^{(2)} \} = \emptyset \, $ as to {\it 
disconnected subsets} if none of the matrix of the first subset $ \, Q_{ \{ \alpha^{(1)} 
\} } \, $ is connected to any matrix of the second one $ \, Q_{ \{ \alpha^{(2)} \} } $.
If one or two external indices belong to a connected subset ($ \, p \in \{ \alpha' \} \, $ or
$ \, q \in \{ \alpha' \} \, $ or $ \, p,q \in \{ \alpha' \} \, $) then all matrices of this subset
are connected to the supermatrices with external indices. These connections break
the supersymmetry for all supermatrices $ \, Q_{ \{ \alpha' \} } $. 

If $ \, m = 2, 3 \, $ then there are no disconnected subsets: all matrices entering 
$ \, {\cal V}^{( m)}_{\{ \alpha \}} \, $ are connected to each other and hence all of them 
are connected to the supermatrices with external indices and the supersymmetry is broken
for all of them.

Starting from $ \, m \ge 4 $, the disconnected subsets may appear and there are $ \, {\cal V
}^{(m)}_{\{ \alpha \}} \, $ containing supermatrices, which are not connected
to any supermatrix with an external index. These are counterparts of the vacuum diagrams 
in the standard interaction representation \cite{AGD} and they {\it do not contribute to the
correlation functions}. For example: (i) if $ \, \{ \alpha^{(1,2)} \} \,  $ are two disconnected 
subsets and $ \, p,q \in \{ \alpha^{(1)} \} \, $ then the supersymmetry is unbroken for
all matrices $ \, \{ Q_{  \{ \alpha^{(2)} \} } \} $, the integrals over the supermatrices
with unbroken supersymmetry yield trivial boundary contributions governed by $ \, Q_{  
\{ \alpha^{(2)} \} }  = 0 \, $ and the corresponding Mayer's functions are zero;
(ii) if $ \, \{ \alpha^{(1,2)} \} \,  $ are two disconnected subsets, $ \, p \ne q \, $ and
$ \, p \in \{ \alpha^{(1)} \} $, $ \, q \in \{ \alpha^{(2)} \} $ then the contribution $ \, 
{\cal V}^{(m)}_{\{ \alpha \}} \, $ is canceled by subtracting the decoupled term
(see the definition of $ \, \langle \langle \ldots \rangle \rangle $).

To summarize the above discussion, each term $ \, {\cal V}^{( m)} \, $ effectively 
consists only of irreducible parts where there are no disconnected matrices. The 
supersymmetry is broken for {\it all} $ \, m \, $ matrices  (i) by the factor $ {\cal R} \, 
{\cal A} $ for the supermatrices $ \, Q_{p,q} \, $ and (ii) by the inter-matrix links for 
the rest supermatrices. To formulate the VE, we focus just on these supermatrices 
$ \, Q_{\alpha_1} \, , \  Q_{\alpha_2} \, \ldots Q_{\alpha_m} $ with broken supersymmetry. 
Let us scale them by the constant $ \, {\cal B} \, $ [\onlinecite{ScalingComm}]:
\begin{equation}\label{Scaling}
   \tilde{Q}_{\alpha_j} = {\cal B} \, Q_{\alpha_j} \, , \quad 1 \le j \le m \, .
\end{equation}
After this scaling, the small parameter $ \, {\cal B} \, $ is eliminated from  $ \, {\cal V}^{( m)} 
\, $ but appears in the single-matrix part $ \, S_0 \, $ in three different ratios as $ \, 1/ {\cal 
B}^2 $, $ \, E / {\cal B} $, and $ \, \Omega / {\cal B} $:
\begin{equation}\label{ScaledS0}
   S_0 \left[ \frac{\tilde{Q}_{\alpha_j}}{ {\cal B} } \right] =
                          {\rm Str} \left\{
      - \frac{1}{4 \, {\cal B}^2 } \, \tilde{Q}_{\alpha_j}^2 + 
            \imath \left( 
                  \frac{E}{ {\cal B} } + 
                  \frac{\Omega}{ {\cal B} } \frac{\Lambda}{2} 
                        \right) \tilde{Q}_{\alpha_j}
                                          \right\} \, .
\end{equation}
The first ratio is large, $ \, 1/ {\cal B}^2  \gg 1 $, and, therefore, the exponentials
\[
   \exp \left\{
          S_0 \left[ \frac{\tilde{Q}_{\alpha_j}}{ {\cal B} } \right] 
            \right\} \propto 
  \exp \left\{ 
              - \frac{1}{4 \, {\cal B}^2 } \, {\rm Str} \left[ \tilde{Q}_{\alpha_j}^2 \right]
           \right\}
\]
suppress the volume of integration over $ \, \tilde{Q}_{\alpha_j} \, $ by
constraint $ \,  {\rm Str} \left[ \tilde{Q}_{\alpha_j}^2 \right] < {\cal B}^2 \, $. We can
draw a conclusion that the larger number $ \, m \, $ of independent $ \, Q $-matrices the smaller 
is the contribution $ \, {\cal G }_{pq}^{(m)}\, $ of $ \, m $-matrix term $ \, {\cal V}^{(m)} 
\, $ to the correlation function. This statement would be unquestionable if one integrates
only over commuting variables. In the case of superintegrals it is more subtle and requires
an additional discussion. An obvious  counterexample to our estimate is the case of
a partition function, for which the supersymmetry is unbroken, $ \, {\cal R} = {\cal A} = 1 $, and 
all superintegrals yield unity regardless of the apparent small phase volume of 
the integration. This happens due to the anomalous contributions  \cite{Efetov} to the superintegrals.
The irreducible terms  $ \, {\cal V}^{( m)}_{\{ \alpha \}} \, $ can also contain the anomalous parts.
The detailed analysis of the correlation function (see Section \ref{VE-verific}) shows
however that, due to the broken supersymmetry, the latter anomalies do not change the 
powers of the small parameter $ \, {\cal B} \, $ and the following estimate holds true:
\begin{equation}\label{VirCoefRat}
   \frac{ {\cal G}_{pq}^{(m+1)} }{ {\cal G}_{pq}^{(m)} } \propto {\cal B} \, , \quad m \ge 2 \, .
\end{equation}
This justifies the supersymmetric virial expansion for the almost diagonal
RMT: {\it the expansion of the correlation function in the number of the interacting supermatrices, 
Eq.(\ref{RA-VE}), yields a regular perturbation theory in powers of $ \, {\cal B} $}.

\section{
       General expression for the virial coefficients
              }\label{SectVirCoef}

\subsection{Saddle-point integration}\label{SaddlIntSect}

Before presenting an expression for the arbitrary number of the interacting matrices,
let us analyze the 2-matrix term $ \, \langle \langle {\cal G}_{pq}^{(2)} \rangle \rangle $
in more detail. As we have already mentioned the supermatrices contained in $ \, 
{\cal V}^{(2)} \, $ must be coupled to the factors $ \, {\cal R}_{p} \, $ and $ \, {\cal 
A}_{p,q} $. Therefore, we obtain the following expression for the off-diagonal 
$ \, p \ne q $ and the diagonal $ \, p = q $ parts of $ \, \langle \langle {\cal G}_{pq}^{(2)} 
\rangle \rangle $:
\begin{eqnarray}
\label{G2}
 \langle \langle
   {\cal G}_{p \ne q}^{(2)}
 \rangle \rangle & = & - \frac{2 \pi}{\Delta} \,
   \Bigl\langle
                  \delta\left( {\rm Str}[ Q_p + Q_q ] \right) \, 
                 {\cal R}_{p} \, {\cal A}_{q} \, {\cal V}^{(2)}_{pq} 
   \Bigr\rangle_{ Q_p, Q_q } \, , \\
\label{G2-diag}
 \langle \langle
   {\cal G}_{pp}^{(2)}
 \rangle \rangle & = & - \frac{2 \pi}{\Delta} \,
   \sum_{n = 1, n \ne p}^N
   \Bigl\langle
                  \delta\left( {\rm Str}[ Q_p + Q_n ] \right) \,
                 {\cal R}_{p} \, {\cal A}_{p} \, {\cal V}^{(2)}_{pn}
   \Bigr\rangle_{ Q_p, Q_n } \, .
\end{eqnarray}
We have introduced the averaging over the supermatrices:
\begin{equation}\label{MatrAver}
   \Bigl\langle \ldots \Bigr\rangle_{Q_{\alpha}} \equiv 
       \int {\cal D} \{ Q_{\alpha} \} ( \ldots ) e^{\tilde{S}_0[Q_{\alpha}]} \, ; \quad 
       \tilde{S}_0[Q_{\alpha}] \equiv S_0[Q_{\alpha}] \Bigl|_{E=0} =
       - \frac{{\rm Str} \left[ Q_{\alpha}^2 \right] }{4} + \imath \frac{\Omega}{2}
                    {\rm Str} \left[ \Lambda \, Q_{\alpha} \right] \, .
\end{equation}
The $ \, \delta $-function  in Eqs.(\ref{G2}-\ref{G2-diag}) resulted from the integral 
over $ \, E $, see Eq.(\ref{S0-def},\ref{RA-VE}). 

Formulas for $ \, {\cal G}_{pq}^{(2)} \, $  exactly describe the contribution of the 
two-level interaction to the correlation function $ \, {\cal G}_{pq} $. However, $ \, 
{\cal V}^{(2)} \, $ contains a product of two supermatrices, cf. Eq.(\ref{Vir2}), which 
entangles the integration variables in a nontrivial way (see Eq.(\ref{Str-Prod})).
We will calculate the integral over the variables $ \, R_{\alpha} $ (see  Eq.(\ref{Str-Terms-1})) 
in the saddle-point approximation.  To explain this step, it is convenient to scale the supermatrices 
by $ \, \sqrt{b_{p \alpha }} $
\[
   \bar{Q}_p = \sqrt{b_{p \alpha }} \, Q_p \, , \quad 
   \bar{Q}_{\alpha} = \sqrt{b_{p \alpha }} \, Q_{\alpha} \, ;
\]
consider the integration over the scaled variables $ \, \bar{R}_{\alpha} \, $ and  $ \, 
\bar{S}_{\alpha} $ in Eqs.(\ref{G2},\ref{G2-diag}) and perform the approximate integration 
over $ \, {\bar R} $:
\begin{eqnarray}
   & & \!\!\!\!\!\!
         { \displaystyle \int \!\!\!\! \int_{-\infty}^{\infty} \!\!\! {\rm d} \bar{R}_{p,\alpha} }
         { \displaystyle \int \!\!\!\! \int_{|\bar{R}_{p,\alpha}|}^{\infty} \!\!\! {\rm d} \bar{S}_{p,\alpha} } \, 
        \delta \! \left( \frac{\bar{R}_p + \bar{R}_{\alpha}}{\sqrt{b_{p \alpha }}} \right) 
    \frac{{\cal R}_p \, e^{- \frac{\bar{R}_p^2 }{4b_{p \alpha } }} }
           {\bar{S}^2_p - \bar{R}^2_p} 
    \frac{{\cal A}_{\alpha} \, e^{- \frac{\bar{R}_{\alpha}^2}{4b_{p \alpha } }} }
           {\bar{S}^2_{\alpha} - \bar{R}^2_{\alpha}} \,
        e^{
             \frac{\imath}{2} \frac{\Omega}{\sqrt{b_{p \alpha }}} ( \bar{S}_p + \bar{S}_{\alpha} )
               }
        {\cal V}^{(2)}_{p,\alpha} \! \left( \bar{R}, \bar{S} \right) \, \simeq \cr
                                & & \label{S-R-int} \\
 & \simeq & 
         \sqrt{ 2 \pi } \, b_{p \alpha } \Biggl(
    { \displaystyle \int \!\!\!\! \int_{0}^{\infty}
            \frac{ {\rm d} \bar{S}_{p,\alpha} }{ \bar{S}^2_p \, \bar{S}^2_{\alpha} }
    }
     {\cal R}_p \, {\cal A}_{\alpha} \Bigl|_{\bar{R}=0} \,
        e^{
             \frac{\imath}{2} \frac{\Omega}{\sqrt{b_{p \alpha }}} 
                          ( \bar{S}_p + \bar{S}_{\alpha} )
               } \, 
        {\cal V}^{(2)}_{p,\alpha} \! \left( 0, \bar{S} \right) + O( \sqrt{b_{p \alpha}})
              + \, O \left( \frac{\bar{R}_t}{\bar{S}_t} \right)
                                                             \Biggr) \, ;
\nonumber
\end{eqnarray}
\[
  \bar{R}_{\alpha} = {\rm Str} [ \bar{Q}_{\alpha} ] \, , \  
  \bar{S}_{\alpha} = {\rm Str} [ \Lambda \, \bar{Q}_{\alpha} ] \, .
\]
Here $ \, \alpha = n, p \, $ for the diagonal- and off-diagonal parts of $ \, {\cal G}^{(2)} $,
respectively. We have accounted for the $ \, \delta $-function in Eqs.(\ref{G2},\ref{G2-diag})
and denoted the typical values of $ \, {\bar S} \, $ and $ \, {\bar R} $, at which the integrals 
converge,  by $ \, {\bar S}_t \, $ and $ \, {\bar R}_t \, $ accordingly. The value of $ \, {\bar R}_t 
\, $ is fixed by the Gaussian exponentials in Eq.(\ref{S-R-int}): $ \, {\bar R}_t \sim \sqrt{
b_{p \alpha}} \, $. The integrals over $ \, {\bar S} \, $ converge due to the exponentials
$ \, e^{ \frac{\imath}{2} \frac{\Omega}{\sqrt{b_{p \alpha }}} \bar{S}_{p,\alpha} } $,
cf. detailed calculations in the Section \ref{Sect2and3Q}. Therefore, we can estimate the
second characteristic scale as $ \, {\bar S}_t \sim \sqrt{b_{p \alpha}}/\omega \, $ 
[\onlinecite{Styp}] and obtain $ \, \bar{R}_t / \bar{S}_t \sim \omega $. The saddle-point
integration over $ \, \bar{R} \, $ makes sense only if the corrections in the right-hand part 
of Eq.(\ref{S-R-int}) are small \cite{SaddleCorr}:
\begin{equation}\label{ValidEstim}
   {\rm max} \left\{ \sqrt{b_{p \alpha}}, \, \frac{\bar{R}_t}{\bar{S}_t} \right\} \sim
   {\rm max} \left\{ \sqrt{b_{p \alpha}}, \, \omega \right\} \ll 1 \, .
\end{equation}
Thus, we have to restrict ourselves to the region $ \, \omega \ll 1 \, $ where the density of
states of the almost diagonal RMTs is close to the constant \cite{DOS}. The ultrahigh frequencies
$ \, \omega \ge 1 \, $ cannot be considered within the saddle-point integration over $ \, R \, $
and they are beyond  the scope of the present paper. We can return to the unscaled matrices 
$ \, Q \, $ and arrive at the following equation for $ \, {\cal G}^{(2)} \, $:
\begin{eqnarray}
\label{G2-L-scale}
 \langle \langle
   {\cal G}_{p \ne q}^{(2)}
 \rangle \rangle & = & - \frac{ ( 2 \pi )^{3/2}}{\Delta} \,
   \Bigl\langle
                  \delta\left( {\rm Str}[ Q_p ] \right) \, \delta\left( {\rm Str}[ Q_q ] \right) \,
                 {\cal R}_{p} \, {\cal A}_{q} \, {\cal V}^{(2)}_{pq} 
   \Bigr\rangle_{ Q_p, Q_q } + \delta {\cal G}_{p \ne q}^{(2)} \, , \\
\label{G2-diag-L-scale}
 \langle \langle
   {\cal G}_{pp}^{(2)}
 \rangle \rangle & = & - \frac{ ( 2 \pi )^{3/2}}{\Delta} \,
   \sum_{n \ne p}^N
   \Bigl\langle
                  \delta\left( {\rm Str}[ Q_p ] \right) \, \delta\left( {\rm Str}[ Q_n ] \right) \,
                 {\cal R}_{p} \, {\cal A}_{p} \, {\cal V}^{(2)}_{pn}
   \Bigr\rangle_{ Q_p, Q_n } + \delta {\cal G}_{pp}^{(2)} \, .
\end{eqnarray}
To calculate the leading terms, we have effectively replaced the exponentials $ \, 
\exp \left( - R^2  / 4 \right) \, $ in the integrand of (\ref{G2}-\ref{MatrAver}) by the 
$ \, \delta $-functions of $ \, R $:
\begin{eqnarray}\label{ExpToDelta}
 & \displaystyle
     \exp \left( - \frac{1}{4} {\rm Str} \left [ Q_p^2 + Q_{\alpha}^2 \right] \right) 
         \delta \Bigl( {\rm Str} \left [ Q_p + Q_{\alpha} \right] \Bigr) =
    \exp \left( - \frac{1}{4} ( R_p^2 + R_{\alpha}^2 ) \right) 
         \delta \left( R_p + R_{\alpha} \right) \to \\
 &    \sqrt{ 2 \pi } \, \delta( R_{\alpha} ) \, \delta( R_p ) =
        \sqrt{ 2 \pi } \, \delta( {\rm Str}[ Q_{\alpha}] ) \, \delta( {\rm Str}[ Q_p ] ) \, .
\nonumber
\end{eqnarray}
Eqs. (\ref{S-R-int},\ref{ValidEstim}) implies in the case of the off-diagonal correlation function:
\begin{equation}\label{2Q-OffDiagCorr}
   \frac{\delta {\cal G}_{p \ne q}^{(2)}}{{\cal G}_{p \ne q}^{(2)}} \sim
   {\rm max} \left\{ {\cal B} \, \sqrt{{\cal F}( | p - q | ) }, \, \omega \right\} \, .
\end{equation}
A similar estimate for the diagonal correlation function is more subtle since the
derivation of $ \, {\cal G}_{pp}^{(2)} \, $ involves the summation over the auxiliary
index $ \, n $. Let us assume that this sum in the leading part of $ \, {\cal G}_{pp}^{(2)} 
\, $ as well as in the correction $ \, \delta {\cal G}_{pp}^{(2)} \, $ converges at a 
characteristic scale $ \, |n-p| \sim  X_{c} $, then we can expect that
\begin{equation}\label{2Q-DiagCorr}
   \frac{\delta {\cal G}_{pp}^{(2)}}{{\cal G}_{pp}^{(2)}} \sim
   {\rm max} \left\{ {\cal B} \, \sqrt{{\cal F}( X_c ) }, \, \omega \right\} \, .
\end{equation}
The value of $ \, X_{c} \, $ is, of course, model dependent and varies for different RMTs.

The generalization of Eqs.(\ref{G2-L-scale},\ref{G2-diag-L-scale}) for an 
arbitrary number $ \, m \, $ of the interacting $ \, Q $-matrices reads
\begin{eqnarray}
\label{Gm-L-scale}
 \langle \langle
   {\cal G}_{p \ne q}^{(m)}
 \rangle \rangle & \simeq & \frac{ 2 \pi }{\Delta} \frac{(-2 \sqrt{\pi})^{m-1}}{\sqrt{m}}
   \sum_{ \{ \alpha_j \ne p, q\} }^N
   \Biggl\langle
                  \delta\left( {\rm Str}[ Q_p ] \right) \, \delta\left( {\rm Str}[ Q_q ] \right) \,
                  {\cal R}_{p} \, {\cal A}_{q} \,
                   \times \\
        & &    \left( \prod_j \delta\left( {\rm Str}[ Q_{\alpha_j} ] \right) \right) \, 
                  {\cal V}^{(m)}_{pq \, \alpha_1 \alpha_2 \ldots \alpha_{m-2}} 
   \Biggr\rangle_{ Q_p, Q_q \, Q_{\alpha_1} Q_{\alpha_2} \ldots Q_{\alpha_{m-2}} }
      1 \le j \le m - 2 \, ; \cr
\label{Gm-diag-L-scale}
 \langle \langle
   {\cal G}_{pp}^{(m)}
 \rangle \rangle & \simeq & \frac{ 2 \pi }{\Delta} \frac{(-2 \sqrt{\pi})^{m-1}}{\sqrt{m}}
   \sum_{ \{ \alpha_j \ne p \} }^N
   \Biggl\langle
                  \delta\left( {\rm Str}[ Q_p ] \right) \,
                  {\cal R}_{p} \, {\cal A}_{p} \,
                   \times \\
        & &    \left( \prod_j \delta\left( {\rm Str}[ Q_{\alpha_j} ] \right) \right) \, 
                  {\cal V}^{(m)}_{pq \, \alpha_1 \alpha_2 \ldots \alpha_{m-1}} 
   \Biggr\rangle_{ Q_p, Q_q \, Q_{\alpha_1} Q_{\alpha_2} \ldots Q_{\alpha_{m-1}} }
      1 \le j \le m - 1 \, .
\nonumber
\end{eqnarray}
The summation is performed over ordered indices $ \, 1 \le \alpha_1 < \alpha_2 < \alpha_3 
< \ldots \le N \, $ excluding the external fixed indices $ \, p \, $ and $ \, q $. It is easy to
show (see section \ref{VE-verific}) that these expressions have the following functional dependence on parameters $\Omega$ and
 ${\cal B}$:
\begin{equation}\label{VE-coef-def}
\langle\langle {\cal G}^{(m)} \rangle\rangle={\cal B}^{m-2}
   \bar{{\cal G}}^{(m)} \left( \Omega / {\cal B} \right),
\end{equation}
with function $\bar{{\cal G}}^{(m)}$ depending only on the ratio $\Omega / {\cal B}$. Thus
one can write down VE of the correlation function as a functional series in powers of $ \, {\cal B} $:
\begin{equation}\label{FuncSer}
   \langle\langle {\cal G}_{pq}({\cal B}, \Omega/{\cal B}) \rangle\rangle
         \simeq \bar{{\cal G}}^D_{pq} + 
         \sum_{m \ge 2} \, {\cal B}^{m-2} \, \bar{{\cal G}}^{(m)}_{pq} (\Omega/{\cal B}) \, .
\end{equation}
The functions $ \, \bar{{\cal G}}^{(m)}_{pq} \, $ are the {\it virial coefficients}. Each 
coefficient $ \, \bar{{\cal G}}^{(m)} (\Omega/{\cal B}) \, $ is governed by the interaction 
of $ \, m \, $ supermatrices corresponding to the interaction of $ \, m \, $ localized states. 
The first term $ \, \bar{{\cal G}}^D \, $ is related to uncorrelated statistics of the diagonal 
part of the almost diagonal RMTs.

By analogy with the estimates (\ref{2Q-OffDiagCorr},\ref{2Q-DiagCorr}), 
Eqs.(\ref{Gm-L-scale},\ref{Gm-diag-L-scale}) describe an arbitrary virial coefficient $ \, 
\bar{{\cal G}}^{(m)} \, $ with the following accuracy:
\begin{equation}\label{mQ-Corr}
   \frac{\delta {\cal G}_{p \ne q}^{(m)}}{{\cal G}_{p \ne q}^{(m)}} \sim
   {\rm max} \left\{ {\cal B} \, \sqrt{{\cal F}( | p - q | ) }, \, \omega \right\} ,
       \quad
   \frac{\delta {\cal G}_{pp}^{(m)}}{{\cal G}_{pp}^{(m)}} \sim
   {\rm max} \left\{ {\cal B} \, \sqrt{{\cal F}( X_c ) }, \, \omega \right\} .
\end{equation}
It is convenient to represent the corrections schematically as a sum of two terms: 
\[
    \delta {\cal G}_{pq}^{(m)} = \delta_{\omega} {\cal G}_{pq}^{(m)} + 
                                                           \delta_{{\cal B}} {\cal G}_{pq}^{(m)}
\]
where $ \, \delta_{\omega} {\cal G}_{pq}^{(m)} / {\cal G
}_{pq}^{(m)} \sim \omega $, $ \,  \delta_{{\cal B}} {\cal G}_{p \ne q}^{(m)}
/ {\cal G}_{p \ne q}^{(m)} \sim {\cal B} \sqrt{{\cal F}( | p - q | )} $, and $ \,  \delta_{{\cal 
B}} {\cal G}_{pp}^{(m)} / {\cal G}_{pp}^{(m)} \sim {\cal B} \sqrt{{\cal F}( X_c )} $.

\subsection{Validity of the virial expansion and the large scale limit}
\label{ValiditySect}

All details of the evaluation of $ \,  {\cal G}^{(2,3)} \, $ and are presented in the next 
section. Here, we would like to discuss the validity and the applicability of VE 
(\ref{Gm-L-scale}-\ref{FuncSer}). 

Firstly we note that VE (\ref{FuncSer}) is a functional series. Its  successive terms
decrease with increasing the number of the interacting supermatrices only if the
absolute value of the virial coefficients is bounded for the arbitrary ratio $ \, \Omega/
{\cal B} $. This condition determines a convergence of VE but it cannot be checked
until the RMT model is specified. In particular, this condition is violated for RMTs
with almost ergodic wavefunctions.

Secondly we should recall that, calculating  $ \, \bar{{\cal G}}^{(m)} 
\left( \Omega / {\cal B} \right) \, $ by the saddle-point approximation in $ \, R $-variables, 
we have neglected corrections $ \,  \delta_{\omega} {\cal G}_{pq}^{(m)} \, $ which 
are of the order of $ \, O(\omega) $. This means that, for a given $\omega$ the 
summation over $ \, m \, $ in VE described by Eqs.(\ref{Gm-L-scale}-\ref{FuncSer}) 
must be stopped at 
\[
    m_{max} \sim 1 + \log(\omega)/\log( {\cal B} ) \, . 
\]
For instance, if $ \, {\cal B} \le \omega \ll 1 \, $ then $ \, \delta_{\omega} 
{\cal G}^{(2)} \ge {\cal G}^{(3)} \, $ and one may take into account only the 
interaction of 2 supermatrices neglecting all higher terms. The next term of VE governed 
by the interaction of  3 supermatrices may be taken into consideration only for the 
smaller energy $ \, \omega \ll  {\cal B} \, $ when $ \, \delta_{\omega} {\cal G}^{(2)} 
\ll {\cal G}^{(3)} $, etc. On the other hand, the neglected dependence of the virial 
coefficients on $ \, \omega \, $ often results from the energy dependence of the density 
of state and does not influence universal properties of the correlation functions.

Finally let us compare $ \, \delta_{\cal B} {\cal G}^{(m)} \, $ 
with the successive term of VE $ \, {\cal G}^{(m+1)} $.  For the sake of simplicity 
we compare $ \, \delta_{\cal B} {\cal G}_{p \ne q}^{(2)} \, $  with $ \, {\cal 
G }_{p \ne q}^{(3)} \, $ though the same analysis applies to the diagonal virial 
coefficients with $ \, p = q \, $ and for the arbitrary $ \, m $. Without loss of generality 
we put $ \, {\cal F}(1) = 1 $. If $ \, | p - q | \sim 1 \, $ then $ \, \delta_{\cal B} {\cal 
G}_{p \ne q}^{(2)} / {\cal G }_{p \ne q}^{(2)} \sim {\cal B} $ and there is no way to 
get the scale separation: $ \, \delta_{\cal B} {\cal G}^{(2)} \, $ and $ \, {\cal G}^{(3)} 
\, $ are of the same order and, again, one may consider only the two matrix
interaction regardless of the energy smallness. 

Now, we will show that $ \, \delta_{\cal B} {\cal G}^{(2)} \, $ is parametrically smaller 
than the higher terms of VE in {\it the large scale limit}. The large scale limit means that 
we consider only those correlation functions which are not sensitive to the details of $ \, 
{\cal F} \, $  at small distances and governed by the behavior of this function  at large 
distances $ \, X_{c} \, $, at which $ \, {\cal F} \, $ is sufficiently small $ \, {\cal F}( X_{c}) 
\ll 1$. It means, in particular, that we assume: (i) $ \, |p-q|\gtrsim X_{c} $ in the case of 
$ \, {\cal G}_{p \ne q}^{(2)} $; (ii) the main contribution to the sum over the auxiliary
index $ \, \alpha_1 \, $ originates from  $ \, |\alpha_1-p| \gtrsim  X_{c} \, $ and $ \, 
|\alpha_1-q| \gtrsim  X_{c} \, $ in the case of $ \, {\cal G}_{p \ne q}^{(3)} $. We 
remind that the sum over $ \, \alpha_1 \, $  is due to presence of the third supermatrix, 
see Eq.(\ref{Gm-L-scale}). Although the second assumption is not applicable for arbitrary 
function  $ \, {\cal F} \, $, it allows to study a wide class of almost diagonal RMTs.
We arrive at the following estimate in the framework of the large scale limit:
\[
   \frac{ \delta_{\cal B} {\cal G}_{p \ne q}^{(2)} }
          { {\cal G }_{p \ne q}^{(2)} } \sim{\cal B} {\cal F}(|p-q|) \le
             {\cal B} {\cal F}( X_{c})  \ll {\cal B} \, .
\] 
The ratio $ \, {\cal G}_{p \ne q}^{(3)} / {\cal G}_{p \ne q}^{(2)} \, $ requires a separate 
consideration: the presence of the third supermatrix in the expression for $ \, {\cal G}_{p \ne q}^{(3)} \,$ results in the additional factors  $ \, {\cal B} \sqrt{ {\cal F} (|p-\alpha_1|)} 
\, $ or $ \, {\cal B} \sqrt{ {\cal F} (|q-\alpha|)} \, $, and simultaneously requires an additional 
summation over the index $ \, \alpha_1 $, see detailed calculations in Sect.\ref{3Col} below. This 
summation is crucial: we have assumed that it converges at $ \, |p-m| \sim |q-m| \sim X_{c} \, $, 
but the large phase volume of summation can {\it compensate the smallness} of $ \, \sqrt{ {\cal F} 
(X_{c})} $, cf.[\onlinecite{Chi-Virial}]. If this is the case we obtain
\[
   \frac{ {\cal G}_{p \ne q}^{(3)} }{ {\cal G}_{p \ne q}^{(2)} } \sim {\cal B} 
      \ \Rightarrow \ 
  \frac{ \delta_{\cal B} {\cal G}_{p \ne q}^{(2)} }{ {\cal G}_{p \ne q}^{(3)} }
     \sim \sqrt{ {\cal F}(X_{c})} \ll 1 \, .
\]
Thus, if the assumptions of the large scale limit hold true then  
$ \, \delta_{\cal B} {\cal G}_{p \ne q}^{(2)} \, $ is parametrically smaller than
 the next term of the virial expansion. In particular if the characteristic scale 
$ \, X_{c} \, $ depends on $ \, N \, $ and the function $ \, {\cal F} \, $ decreases 
in such a way  that $ \, \lim_{N \to \infty} \sqrt{ {\cal F} (X_{c}) } = 
0 \, $ then the ratio $ \, \delta_{\cal B} {\cal G}_{p \ne q}^{(2)} / {\cal G }_{p \ne q}^{(2,3)} 
\, $ asymptotically goes to zero. This scale separation justifies VE in the large scale limit 
in many cases. One can check, for example,  that it is correct for the spectral statistics of the 
critical almost diagonal PLBRMs, where  $ \, X_{c} \propto {\cal B } N \, $ and the relevant 
energy range reads $ \, \omega \le {\cal B} \Delta $ [\onlinecite{Trotter}].

We would like to mention that the large scale limit considered here is analogous to the diffusive
approximation of the standard $ \, \sigma $-model. In the latter approximation the spatial 
scales large compared to the mean free path are assumed to be the only relevant ones. In the 
same time the saddle-point approximation justified in the large scale limit in our approach results 
in the linear constraint on the $ \, Q $-matrix
\begin{equation}
   {\rm Str}[Q] = 0,
\end{equation} 
while the saddle-point manifold of the standard diffusive $ \, \sigma $-model is defined by the
additional nonlinear constraint $ \, \bigl( Q^{(\sigma)} \bigr)^2 = 1 $.

\subsection{Correlation function at the band center}\label{BandCenterCorr}

The expressions for the correlation functions given by Eq.(\ref{Gm-L-scale}) 
and Eq.(\ref{Gm-diag-L-scale}) were derived after integration over $E$. However
in some applications it is more convenient to consider a correlation function 
at a fixed energy. The aim of this  section is to discuss briefly the correlation
functions at the band center $E=0$.

We define the two-point correlation function at the band center as
\begin{equation}\label{Gpq-cent}
    \bar{\cal G}_{pq}( \omega, E=0 ) \equiv 
         \hat{G}^R_{pp} (\omega/2 ) \hat{G}^A_{qq} ( - \omega/2 ) \, .
\end{equation}
Repeating all the steps leading to the results Eq.(\ref{Gm-L-scale},\ref{Gm-diag-L-scale}),
we obtain:
\begin{equation}
\label{Gm-Cent}
 \langle \langle \bar{\cal G}^{(m)}_{pq}( \omega, E=0 ) \rangle \rangle \simeq
      \frac{\sqrt{m}}{N}
      \langle \langle {\cal G}^{(m)}_{pq}( \omega) \rangle \rangle \, .
\end{equation}
The ratio $ \, \sqrt{m} / N \, $ can be referred to as ``the unfolding factor'' \cite{Errat}.

Let us also note that an average of the product of two retarded (advanced) Green's
functions can be neglected again for the following reason. If we look at the parametrization
of $Q$ in this case (Appendix \ref{QRR-param}), then we notice that variables $R$ and
$S$ change their roles. For this reason the large scale approximation ${\rm Str}[Q]\approx 0$ 
implies now $\lambda_R,\,\lambda_{R'}\approx 0$. Thus the volume of the integration in 
this case becomes parametrically small at $ \, {\cal B} \ll 1 \, $ and $ \, \omega \ll 1 $.

Below, we will analyze only the correlation functions averaged over $ \, E $.

\section{The cases of 2- and 3-matrix approximation}\label{Sect2and3Q}

In this section we present detailed calculation of the contributions $ \, {\cal G}^{(2)} 
\, $ and $ \, {\cal G}^{(3)} \, $ governed by the interaction of 2 and 3 supermatrices
respectively. It is more convenient to expand $ \, {\cal V}^{(2,3)} \, $ in the series
of powers of $ \, Q $-matrices and then integrate over $Q$ term by term. 
This step is not essential for the approximation of two interacting matrices which can be
worked out directly from Eqs.(\ref{G2-L-scale},\ref{G2-diag-L-scale}).
However, it is more convenient for the 3-matrix approximation, since
it allows one to unify the calculations for the different number of
the interacting supermatrices 
and to avoid an explicit derivation of anomalous terms in the
superintegrals. The disadvantage of this route is that the series obtained 
converge only asymptotically and one has to Fourier transform them in
order to analyze  the answer in the time-domain \cite{Trotter}.

\subsection{The case of 2 interacting supermatrices}\label{2Col}

We start with calculating $ \, {\cal G}^{(2)} \, $ using approximate
formulas (\ref{G2-L-scale},\ref{G2-diag-L-scale}). The power series 
for $ \, {\cal V}^{(2)} \, $ reads:
\begin{equation}
    {\cal V}^{(2)}_{pn} = \sum_{k=1}^{\infty} 
                          \frac{ \left( -2 b_{pn} {\rm Str} [ Q_p Q_n ] \right)^k }{k!} \, .
\end{equation}
We use the phase $ \, \phi \, $ and two non-compact variables $ \, R \, $ and $ \, S \, $
to parameterize the boson-boson sector of each supermatrix  (see the corresponding 
definitions in Appendix \ref{Q-param}). The integration measure in 
Eqs.(\ref{G2-L-scale},\ref{G2-diag-L-scale}) takes the form:
\begin{equation}\label{Q-meas}
        \int {\cal D} \{ Q \}\delta ( {\rm Str}[Q] )  \Bigl( \ldots \Bigr) \to 2
        \int_{-\infty}^{\infty} \!\!\! {\rm d} R  \, \delta(R) \, \int_0^{\infty} \frac{  {\rm d} S }{ S^2 } \ 
                 \frac{1}{2\pi} \, \int_0^{2 \pi} \!\!\! {\rm d} \phi  \ 
                 \int {\rm d} \{ \eta^*_R \eta_R \eta^*_A \eta_A \} \Bigl( \ldots \Bigr) \,.
\end{equation}
and the expressions for $ \, {\cal R}_p  {\cal A}_{q,p} \, $ read:
\begin{eqnarray}
\label{PE-2Col-OffD}
     {\cal R}_p  {\cal A}_q \Bigl|_{R_{p,q}=0} & = & \frac{1}{4} S_p S_q \ 
              \bigl( \eta^*_R \eta_R \bigr)_p \bigl( \eta^*_A \eta_A \bigr)_q \,  ;
  \\
\label{PE-2Col-D}
     {\cal R}_p  {\cal A}_p \Bigl|_{R_p =0} & = & \frac{1}{4} S_p^2 \ 
              \bigl( \eta^*_R \eta_R \eta^*_A \eta_A \bigr)_p \, .
\end{eqnarray}
Integrating over $ \, R $-variables we obtain
\begin{eqnarray}
\label{G2-2Col}
 \langle \langle
   {\cal G}_{p \ne q}^{(2)}
 \rangle \rangle & \simeq & - \frac{ ( 2 \pi )^{3/2}}{\Delta} \, \sum_{k=1}^{\infty} 
         \int \!\!\!\! \int_0^{\infty} \frac{ {\rm d} S_{p,q} }{ S_p S_q } \,
                  e^{\imath \frac{\Omega}{2} ( S_p + S_q )} \,
         \int {\rm d} \{ \eta^*_R \eta_R \eta^*_A \eta_A \}_{p,q} \times \cr
  & & \quad \times
            \bigl( \eta^*_R \eta_R \bigr)_p \bigl( \eta^*_A \eta_A \bigr)_q \,
         \int \!\!\!\! \int_0^{2\pi} \frac{ {\rm d} \phi_{p,q}}{(2\pi)^2}
                  \frac{ \left( -2 b_{pq} \, {\cal S}_{pq} \right)^k }{k!} \, , \\
\label{G2-diag-2Col}
 \langle \langle
   {\cal G}_{pp}^{(2)}
 \rangle \rangle & \simeq & - \frac{ ( 2 \pi )^{3/2}}{\Delta} \,
   \sum_{n \ne p}^N \sum_{k=1}^{\infty} 
         \int \!\!\!\! \int_0^{\infty} \frac{ {\rm d} S_{p,n} }{ S_n^2 } \,
                  e^{\imath \frac{\Omega}{2} ( S_p + S_n )} \,
         \int {\rm d} \{ \eta^*_R \eta_R \eta^*_A \eta_A \}_{p,n} \times \cr
  & & \quad \times
            \bigl( \eta^*_R \eta_R \eta^*_A \eta_A \bigr)_p \,
         \int \!\!\!\! \int_0^{2\pi} \frac{ {\rm d} \phi_{p,n}}{(2\pi)^2}
                  \frac{ \left( -2 b_{pn} \, {\cal S}_{pn} \right)^k }{k!} \, .
\end{eqnarray}
The expression for $ \, {\cal S}_{pq} \equiv {\rm Str} [ Q_p Q_q] \Bigl|_{R_{p,q}=0} 
\propto S_p S_q \, $ is given in Appendix \ref{PhAver}, Eqs.(\ref{Av-2},\ref{Av-4}). The 
integrals of $ \, {\cal S}_{pq}^k \, $ over the phases are calculated in the same Appendix, 
Eqs.(\ref{OnePhAver},\ref{Av-3}). The integrals over the $ \, S $-variables are 
regularized at the upper limit 
by the imaginary part of $ \, \Omega $ and converge at the lower limit for all $ \, k \, $ in 
the case of $ \, {\cal G}_{p \ne q}^{(2)} \, $ and for $ \, k \ge 2 \, $ in the case of $ \, 
{\cal G}_{pp}^{(2)} $. The term with $ \, k = 1 \, $ in the diagonal part $ \, {\cal 
G}_{pp}^{(2)} $ is special: it is governed by an anomaly, i.e., an uncertainty $ \,
0 \times \infty \, $ with zero resulting from  the integrals over the Grassmann 
variables $ \, \bigl( \eta^*_R \eta_R \eta^*_A \eta_A \bigr)_n \, $ and infinity due
to the divergence at the lower limit of integration over the commuting variable $ \, S_n $. 
This uncertainty can be resolved either in a standard way \cite{Efetov} or, equally,  one 
can calculate the integrals in the diagonal part for $ \, k \ge 2 \, $ and then perform an 
analytic continuation for $ \, k = 1 $. The result of the integration over all variables can 
be written as follows:
\begin{eqnarray}
\label{G2-2Col-ser}
 \langle \langle
   {\cal G}_{p \ne q}^{(2)}
 \rangle \rangle & \simeq & \frac{ (2\pi)^{3/2} }{\Delta} \, \sum_{k=1}^{\infty}
            \left( \frac{2 b_{pq}}{\Omega^2} \right)^k
            \frac{\Gamma(2k-1)}{\Gamma(k)} (k-1) \, , \\
\label{G2-diag-2Col-ser}
 \langle \langle
   {\cal G}_{pp}^{(2)}
 \rangle \rangle & \simeq & \frac{ (2\pi)^{3/2} }{\Delta} \,
   \sum_{n \ne p}^N \sum_{k=1}^{\infty} 
            \left( \frac{2 b_{pn}}{\Omega^2} \right)^k
            \frac{\Gamma(2k-1)}{\Gamma(k)} k \, .
\end{eqnarray}
The correlation functions
$ \, R_2 \, $ and $ \, C_2 \, $ can be calculated from the real part of $ \, {\cal G}_{pq} \, $ (see Eqs.(\ref{R2-int},\ref{C2-int})). Taking the real part by substituting $ \, \omega \, $ 
instead of $ \, \Omega \, $ in Eqs. (\ref{G2-2Col-ser},\ref{G2-diag-2Col-ser}) one obtains the asymptotic series in the energy representation. However it is more convenient to consider the
time representation by performing the Fourier transform of the real part
\[
   {\cal G}_{pq}(t) = \frac{1}{2 \Delta} \int {\rm d} \omega \, e^{- \imath \, \omega t}
           \bigl( {\cal G}_{pq}(\omega) + c.c. \bigr),
\]
obtaining
\begin{eqnarray}
\label{G2-2Col-ser-Four}
 \langle \langle
   {\cal G}_{p \ne q}^{(2)}(t)
 \rangle \rangle & \simeq & \frac{ \pi (2\pi)^{3/2} }{\Delta^2 |t| } \, \sum_{k=1}^{\infty}
            \frac{\left( - 2 b_{pq} t^2 \right)^k}{(k-1)!} \frac{k-1}{2k-1} \, , \\
\label{G2-diag-2Col-ser-Four}
 \langle \langle
   {\cal G}_{pp}^{(2)}(t)
 \rangle \rangle & \simeq & \frac{ \pi (2\pi)^{3/2} }{\Delta^2 |t| } \,
   \sum_{n \ne p}^N \sum_{k=1}^{\infty}             
            \frac{\left( - 2 b_{pn} t^2 \right)^k}{(k-1)!} \frac{k}{2k-1} \, .
\end{eqnarray}
Now the summation over $ \, k \, $ can be done explicitly
\begin{eqnarray}
\label{G2-2Col-Four}
 \langle \langle
   {\cal G}_{p \ne q}^{(2)}(t)
 \rangle \rangle & \simeq & - \sqrt{2} \frac{ \pi^{5/2} }{\Delta^2} \, 
     \sqrt{2 b_{pq}}
        \left[ 
           \sqrt{2 b_{pq}} |t| e^{ - 2 b_{pq} t^2 } -
           \frac{\sqrt{\pi}}{2} {\rm erf} \left( \sqrt{2 b_{pq}} |t| \right)
        \right] \, , \\
\label{G2-diag-2Col-Four}
 \langle \langle
   {\cal G}_{pp}^{(2)}(t)
 \rangle \rangle & \simeq & - \sqrt{2} \frac{ \pi^{5/2} }{\Delta^2} \,
   \sum_{n \ne p}^N 
     \sqrt{2 b_{pn}}
        \left[ 
           \sqrt{2 b_{pn}} |t| e^{ - 2 b_{pn} t^2 } +
           \frac{\sqrt{\pi}}{2} {\rm erf} \left( \sqrt{2 b_{pn}} |t| \right)
        \right] \, .
\end{eqnarray}
Here $ \, {\rm erf} (z) = \frac{2}{\sqrt{\pi}} \int_0^z e^{- t^2} {\rm d} t $. Our 
theory can be verified by comparison with the results of TVE. To this
end  we calculate the form factor, which is the Fourier transform of the two-level 
correlation function $ \, R_2 $:
\begin{equation}\label{FormFactor}
    K(t) = \frac{\Delta^2}{2 \pi^2 N} \Re \sum_{p,q=1}^N 
                   \langle \langle {\cal G}_{pq}(t) \rangle \rangle \, ,
\end{equation}
and insert in this formula Eqs.(\ref{G2-2Col-Four},\ref{G2-diag-2Col-Four}). This
gives the form factor in the approximation of two interacting levels:
\begin{eqnarray}\label{FormFactor-2Col}
    K^{(2)}(t) & \simeq & - \frac{\sqrt{2\pi}}{ N |t| } \sum_{p,q=1}^N 
             x(|p-q|)  e^{ - x(|p-q|)} \simeq \Bigl|_{N \gg 1} 
                                           - 2 \frac{\sqrt{2\pi}}{ |t| } \sum_{m=1}^N 
                                                   x(m)  e^{ - x(m)} \, , \\
    x(|p-q|) & \equiv & 2 b_{pq} t^2 = \frac{1}{2} ( {\cal B} t )^2 {\cal F}(|p-q|) \, ;
\nonumber
\end{eqnarray}
which coincides with the expression for $ \, K^{(2)}(t) \, $ obtained by TVE \cite{Trotter}.

This comparison of TVE and the theory based on SuSyFT clearly demonstrates that SuSyFT
is capable to give much more detailed information on the correlation functions. Namely, 
TVE deals with the form factor which is an integral quantity obtained after the summation 
of diagonal and off-diagonal parts of the correlation function $ \, {\cal G} \, $ over all spatial
coordinates, while the correlation function $ \, {\cal G} \, $ at given spatial points can be 
derived only from SuSyFT.

\begin{figure}[t]
\unitlength1cm
\begin{picture}(11.0,8)
   \epsfig{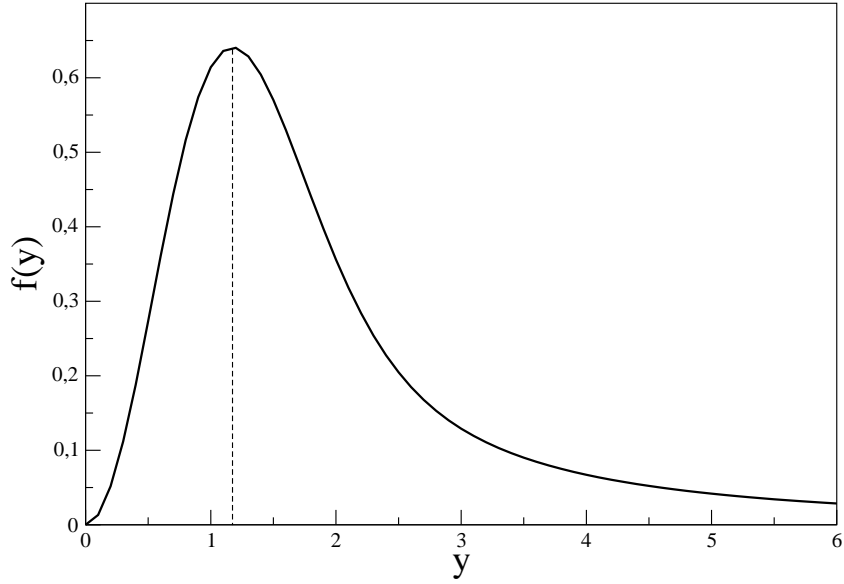}
\end{picture}
\vspace{0.5cm}
\caption{
\label{Summand}
Summand in the right hand side of Eq.(\ref{G2-diag-2Col-final}): $ \,
f(y) = - \left[ 1 - \frac{\sqrt{\pi}}{2} e^{-y^2} \left( 2 y + \frac{1}{y} 
\right) \, {\rm erfi} (y) \, \right]  $. Here the argument $ \, y \, $ denotes
the parameter $ \, \omega / \sqrt{8 b_{pn}} $.
}
\end{figure}

We can now return  from Eqs.(\ref{G2-2Col-Four},\ref{G2-diag-2Col-Four}) written in 
the time-domain to the energy representation of $ \, \Re \left[ {\cal G}^{(2)} \right] $:
\begin{eqnarray}
\label{G2-2Col-final}
 \Re \,
 \langle \langle
   \, {\cal G}_{p \ne q}^{(2)}(\omega) \,
 \rangle \rangle \!\! & \simeq & \!\! - \frac{\pi^{3/2}}{\sqrt{2} \Delta}
        \left[ 
            1 - \frac{ \sqrt{\pi} }{2} e^{-\frac{ \omega^2 }{8 b_{pq}} } \,
               \left( \frac{\omega}{ \sqrt{2 b_{pq}} } - \frac{ \sqrt{8 b_{pq}} }{\omega} \right) \, 
                                {\rm erfi } \left( \frac{ \omega }{ \sqrt{8 b_{pq}} } \right)
        \right] \! , \\
\label{G2-diag-2Col-final}
 \Re \,
 \langle \langle
  \, {\cal G}_{pp}^{(2)}(\omega) \,
 \rangle \rangle \!\! & \simeq & \!\! - \frac{\pi^{3/2}}{\sqrt{2} \Delta}
   \sum_{n \ne p}^N 
        \left[ 
            1 - \frac{ \sqrt{\pi} }{2} e^{-\frac{ \omega^2 }{8 b_{pn}} } \,
               \left( \frac{\omega}{ \sqrt{2 b_{pn}} } + \frac{ \sqrt{8 b_{pn}} }{\omega} \right) \, 
                                {\rm erfi } \left( \frac{ \omega }{ \sqrt{8 b_{pn}} } \right)
        \right] \! .
\end{eqnarray}
The power series Eqs.(\ref{G2-2Col-ser},\ref{G2-diag-2Col-ser}) are asymptotic
expansion of these formulas \cite{AbrSteg}. Note that the summands in the right hand 
side of Eq.(\ref{G2-diag-2Col-final}) are peaked around the value
\begin{equation} 
   \frac{\omega}{\sqrt{8 b_{pn}}} \equiv 
   \frac{\omega}{2 {\cal B} \sqrt{{\cal F}(|p-n|)}}  \sim 1 \, , 
\end{equation}
see Fig.\ref{Summand}. Thus, we can find the characteristic spatial scale $ \, X_{c} $, 
which yields the main contribution to the sum over $ \, n \, $ and determines $ \, {\cal 
G}_{pp}^{(2)}(s) $, from the following estimate
\begin{equation}\label{Scale-2Q}
  {\cal F}(X_{c}) \sim \left( \frac{\omega}{ {\cal B} } \right)^2 \, .
\end{equation}
Estimate (\ref{Scale-2Q}) ensures the validity of the large scale limit for $ \, 
\omega \ll {\cal B} \ll 1 \, $ at the level of 2 matrix  approximation. Indeed if 
$ \, \omega \ll {\cal B}\, $ then $ \, {\cal F}(X_{c}) \ll 1 \,$, hence $\, X_c \gg 
1 $, i.e., the diagonal correlator $ \, {\cal G}^{(2)}_{pp} \, $ is governed by 
the large distances, and the correction $ \, \delta_{{\cal B}} \, {\cal 
G}^{(2)} \, $ to the saddle-point integration is expected to be smaller than 
the higher terms of the VE. On the contrary, in the range $ \, {\cal B} \le \omega 
\ll 1 \, $ the characteristic scale is small, $ \, X_c \to 1 \, $ and the higher terms of 
the VE can be of the same order as the omitted correction $ \, \delta_{{\cal B}} \, 
{\cal G}^{(2)} $, see the Section \ref{ValiditySect}. This means that we 
cannot use the saddle-point integration to go beyond the two-matrix approximation
in the case $ \, {\cal B} \le \omega \ll 1 $.

\subsection{The case of 3 interacting supermatrices}\label{3Col}

The calculations of $ \, {\cal G}^{(3)} \, $ based on Eqs.(\ref{Gm-L-scale},\ref{Gm-diag-L-scale}) 
are very similar to those described in the preceding section for $ \,  {\cal G}^{(2)} $. The power 
series for $ \, {\cal V}^{(3)} \, $ read
\begin{eqnarray}\label{V3-ser}
    {\cal V}^{(3)}_{pmn} & = &
        \Bigl\{ 
             \sum_{k_{1,2,3}=1}^{\infty} + 
             \sum_{k_{1,2}=1}^{\infty}\Bigl|_{k_3=0} +
             \sum_{k_{1,3}=1}^{\infty}\Bigl|_{k_2=0} +
             \sum_{k_{2,3}=1}^{\infty}\Bigl|_{k_1=0}
        \Bigr\} \\
           & &
                          \frac{ \left( -2 b_{pn} {\rm Str} [ Q_p Q_n ] \right)^{k_1} }{k_1!} 
                          \frac{ \left( -2 b_{pm} {\rm Str} [ Q_p Q_m ] \right)^{k_2} }{k_2!}
                          \frac{ \left( -2 b_{mn} {\rm Str} [ Q_m Q_n ] \right)^{k_3} }{k_3!} \, .
\nonumber
\end{eqnarray}
We insert this series in Eqs.(\ref{Gm-L-scale},\ref{Gm-diag-L-scale}) and integrate over 
$ \, R $-variables obtaining
\begin{eqnarray}
\label{G2-3Col}
 \langle \langle
   {\cal G}_{p \ne q}^{(3)}
 \rangle \rangle & \simeq & \frac{ (4 \pi )^2}{\sqrt{3}\Delta} \, 
         \sum_{ \{ m \ne p,q \} }^N \sum_{k_{1,2,3}=0}^{\infty} 
         \int \!\!\!\! \int \!\!\!\! \int_0^{\infty} \frac{ {\rm d} S_{p,q,m} }{ S_p S_q S_m^2} \,
                  e^{\imath \frac{\Omega}{2} ( S_p + S_q + S_m )} \,
         \int {\rm d} \{ \eta^*_R \eta_R \eta^*_A \eta_A \}_{p,q,m} \times \\
   \times & & \!\!\!\!\!\!\!\!
            \bigl( \eta^*_R \eta_R \bigr)_p \bigl( \eta^*_A \eta_A \bigr)_q \,
         \int \!\!\!\! \int \!\!\!\! \int_0^{2\pi} \frac{ {\rm d} \phi_{p,q,m}}{(2\pi)^3}
                  \frac{ \left( -2 b_{pq} {\cal S}_{pq} \right)^{k_1} }{k_1!} 
                  \frac{ \left( -2 b_{pm} {\cal S}_{pm} \right)^{k_2} }{k_2!}
                  \frac{ \left( -2 b_{qm} {\cal S}_{qm} \right)^{k_3} }{k_3!} \, , \cr
   \cr
\label{G2-diag-3Col}
 \langle \langle
   {\cal G}_{pp}^{(3)}
 \rangle \rangle & \simeq & \frac{ (4 \pi )^2}{\sqrt{3}\Delta} \,
   \sum_{ \{m,n \ne p \} }^N \sum_{k_{1,2,3}=0}^{\infty} 
         \int \!\!\!\! \int \!\!\!\! \int_0^{\infty} \frac{ {\rm d} S_{p,m,n} }{ S_m^2 S_n^2 } \,
                  e^{\imath \frac{\Omega}{2} ( S_p + S_m + S_n )} \,
         \int {\rm d} \{ \eta^*_R \eta_R \eta^*_A \eta_A \}_{p,m,n} \times \\
   \times & & \!\!\!\!\!\!\!\!
            \bigl( \eta^*_R \eta_R \eta^*_A \eta_A \bigr)_p \,
         \int \!\!\!\! \int \!\!\!\! \int_0^{2\pi} \frac{ {\rm d} \phi_{p,m,n}}{(2\pi)^3}
                  \frac{ \left( -2 b_{pq} {\cal S}_{pq} \right)^{k_1} }{k_1!} 
                  \frac{ \left( -2 b_{pm} {\cal S}_{pm} \right)^{k_2} }{k_2!}
                  \frac{ \left( -2 b_{qm} {\cal S}_{qm} \right)^{k_3} }{k_3!} \, ; \ m > n \, .
\nonumber
\end{eqnarray}
We use the same trick with the analytical continuation from $ \, k_j \ge 2 \, $ to $ \, k_j = 1 \, $
to handle the anomalous terms. Note that we have put zero in the lower limit for the sum over 
$ \, k_{1,2,3} \, $ and combined all 4 contribution in Eq.(\ref{V3-ser}) together. This is possible 
since all terms with either $ \, k_1 = k_2 = 0 \, $ or $ \, k_1 = k_3 = 0 \, $ or $ \, k_2 = k_3 = 0 \, $ 
are equal to zero after the analytical continuation (see the results for $ \, {\cal G}^{(3)} \, $ below).
The integrals over the phases are calculated in Appendix~\ref{PhAver}, see 
Eqs.(\ref{2PhAver-1}--\ref{2PhAver-8}). These rather cumbersome expressions are substantially
simplified after the integration over the Grassmann variables (see Appendix \ref{ThanksMaple}):
\begin{eqnarray}\label{3Col-Ph-Gr-Int}
& &
  \int {\rm d} \{ \eta^*_R \eta_R \eta^*_A \eta_A \}_{p,q,m}
   \bigl( \eta^*_R \eta_R \bigr)_p \bigl( \eta^*_A \eta_A \bigr)_q \,
         \int \!\!\!\! \int \!\!\!\! \int_0^{2\pi} \frac{ {\rm d} \phi_{p,q,m}}{(2\pi)^3}
                  \left( {\cal S}_{pq} \right)^{k_1} 
                  \left( {\cal S}_{pm} \right)^{k_2}
                  \left( {\cal S}_{qm} \right)^{k_3} = \cr
& = &
  \int {\rm d} \{ \eta^*_R \eta_R \eta^*_A \eta_A \}_{p,q,m}
   \bigl( \eta^*_R \eta_R \eta^*_A \eta_A \bigr)_p \,
         \int \!\!\!\! \int \!\!\!\! \int_0^{2\pi} \frac{ {\rm d} \phi_{p,q,m}}{(2\pi)^3}
                  \left( {\cal S}_{pq} \right)^{k_1} 
                  \left( {\cal S}_{pm} \right)^{k_2}
                  \left( {\cal S}_{qm} \right)^{k_3} = \cr
& = &
     \frac{1}{32} \ \Xi ( k_1, k_2, k_3 ) \ 
     \frac{S_p^{k_1+k_2}}{\Gamma(k_1+k_2 -1)} 
     \frac{S_q^{k_1+k_3}}{\Gamma(k_1+k_3 -1)} 
     \frac{S_m^{k_2+k_3}}{\Gamma(k_2+k_3 -1)} \, ;
\end{eqnarray}
where
\begin{eqnarray}
   \Xi ( k_1, k_2, k_3 ) & = &
              \frac{\Gamma(k_1 - 1/2)}{\pi^{1/2} \, k_1! } 
              \frac{\Gamma(k_2 - 1/2)}{\pi^{1/2} \, k_2! }
              \frac{\Gamma(k_3 - 1/2)}{\pi^{1/2} \, k_3!} \times \\
   & & \quad
                \times ( 2 k_1 k_2 k_3 - k_1 k_2 - k_1 k_3 - k_2 k_3 ) \, \Gamma(k_1 + k_2 +k_3 -1) \, .
\nonumber
\end{eqnarray}
After the integration over $ \, S $-variables, the power series for $ \, {\cal G}^{(3)} \, $ take 
the following form:
\begin{eqnarray}
\label{G2-3Col-ser}
 \langle \langle
   {\cal G}_{p \ne q}^{(3)}
 \rangle \rangle & \simeq & - \frac{\imath \, \pi^2}{4 \sqrt{3}} \, \frac{\Omega}{\Delta} \, 
         \sum_{ \{ m \ne p,q \} }^N \sum_{k_{1,2,3}=0}^{\infty} 
            \left( \frac{ 8 b_{pq} }{\Omega^2} \right)^{k_1}
            \left( \frac{ 8 b_{pm} }{\Omega^2} \right)^{k_2}
            \left( \frac{ 8 b_{qm} }{\Omega^2} \right)^{k_3} \times \\
   & & 
            \times \ \Xi ( k_1, k_2, k_3 ) \ ( k_1 + k_2 - 1 ) \, ( k_1 + k_3 - 1 ) \, , \cr 
    \cr
\label{G2-diag-3Col-ser}
 \langle \langle
   {\cal G}_{pp}^{(3)}
 \rangle \rangle & \simeq & - \frac{\imath \,  \pi^2}{8 \sqrt{3}} \, \frac{\Omega}{\Delta} \,
   \sum_{ \{m,n \ne p \} }^N \sum_{k_{1,2,3}=0}^{\infty} 
            \left( \frac{ 8 b_{pm} }{\Omega^2} \right)^{k_1}
            \left( \frac{ 8 b_{pn} }{\Omega^2} \right)^{k_2}
            \left( \frac{ 8 b_{mn} }{\Omega^2} \right)^{k_3}  \times \\
   & & 
             \times \ \Xi ( k_1, k_2, k_3 ) \ ( k_1 + k_2 ) ( k_1 + k_2 - 1) \, .
\nonumber
\end{eqnarray}
Following the procedure described in the preceding section, we Fourier transform
the real part of Eqs.(\ref{G2-3Col-ser},\ref{G2-diag-3Col-ser}) and obtain:
\begin{eqnarray}
\label{G2-3Col-Four}
 \langle \langle
   {\cal G}_{p \ne q}^{(3)}(t)
 \rangle \rangle & \simeq & \frac{\pi^3}{4 \sqrt{3}} \, \frac{1}{( t \, \Delta )^2} \, 
         \sum_{ \{ m \ne p,q \} }^N \sum_{k_{1,2,3}=0}^{\infty} 
            \left( - 8 b_{pq}  t^2 \right)^{k_1}
            \left( - 8 b_{pm} t^2 \right)^{k_2}
            \left( - 8 b_{qm} t^2 \right)^{k_3} \times \\
   \times & & \!\!\!\!\!\!\!
            \frac{ \Xi ( k_1, k_2, k_3 ) }{ \Gamma\bigl( 2[k_1+k_2+k_3] - 1 \bigr) } \, 
            ( k_1 + k_2 - 1 ) \, ( k_1 + k_3 - 1 ) \, , \cr 
    \cr
\label{G2-diag-3Col-Four}
 \langle \langle
   {\cal G}_{pp}^{(3)}(t)
 \rangle \rangle & \simeq & \frac{\pi^3}{8 \sqrt{3}} \, \frac{1}{( t \, \Delta )^2} \,
   \sum_{ \{m,n \ne p \} }^N \sum_{k_{1,2,3}=0}^{\infty} 
            \left( - 8 b_{pm} t^2 \right)^{k_1}
            \left( - 8 b_{pn} t^2 \right)^{k_2}
            \left( - 8 b_{mn} t^2 \right)^{k_3}  \times \\
   \times & & \!\!\!\!\!\!\!
             \frac{ \Xi ( k_1, k_2, k_3 ) }{ \Gamma\bigl( 2[k_1+k_2+k_3] - 1 \bigr) } \, 
             ( k_1 + k_2 ) ( k_1 + k_2 - 1 ) \, .
\nonumber
\end{eqnarray}
The triple sums  on  the r.h.s. of Eqs.(\ref{G2-3Col-Four},\ref{G2-diag-3Col-Four}) 
 can not be reduced to a product of simple sums. Therefore,
the summation over $ \, k_{1,2,3} \, $ is not trivial \cite{Chi-Virial}. To verify SuSyFT,
we calculate the contribution of three interacting matrices  to the form factor (\ref{FormFactor}).
The answer looks more compact if we at first symmetrize the expression for $ \, {\cal G}^{(3)}
(t) \, $ with respect to $ \, k_{1,2,3} \, $ and then turn to the ordered sum over 3 remaining indices:
\begin{eqnarray}\label{FormFactor-3Col}
   K^{(3)}(t) & \simeq & \frac{2}{\sqrt{3} \, t^2} \frac{1}{N} \,
   \sum_{ \{ m > n > p \} }^N \sum_{k_{1,2,3}=0}^{\infty} 
            \left( - 2 b_{pm} t^2 \right)^{k_1}
            \left( - 2 b_{pn} t^2 \right)^{k_2}
            \left( - 2 b_{mn} t^2 \right)^{k_3}  \times \\
   \times & & \!\!\!\!\!\!\!
             \frac{ 4^{k_1+k_2+k_3-1} \, \pi \ \Xi ( k_1, k_2, k_3 ) }{ \Gamma\bigl( 2[k_1+k_2+k_3] - 1 \bigr) } \, 
             ( k_1 + k_2 + k_3 - 1 ) (k_1 + k_2 + k_3 - 3/2 ) \, .
\nonumber
\end{eqnarray}
Eq.(\ref{FormFactor-3Col}) coincides with the expression for $ \, K^{(3)}(t) \, $ obtained
by TVE \cite{Trotter}. 

We remind that all results of this section should be taken into account in the VE 
if $ \, \omega \ll {\cal B} \ll 1 \, $ and the assumptions of the large scale limit hold 
true, i.e., the sums over $ \, m \, $ and $ \, n \, $ 
converge at the large spatial scales. This is the case, for instance, for the spectral statistics of
the critical PLBRMs where the relevant energy range is small $ \, \omega < {\cal B } \, \Delta 
\, $ while the characteristic spatial scale, which governs the two level correlations in the 
framework of 2- and 3-matrix approximation, is large $ \, X_{c} \sim {\cal B} / \omega 
\gg 1 $ [\onlinecite{Chi-Virial}].

\subsection{Verification of the supersymmetric VE}\label{VE-verific}

Let us verify that the power series obtained for $ \, {\cal G}^{(m)} \, , \ m \ge 2 $, 
really obey the estimate (\ref{VirCoefRat}). If we consider a term of the power
series with given powers $ \, k_j \, $, integrate it over all $ \, R $-variables
in the saddle-point approximation, scale $ \, S $-variables by $ \, \Omega  $, and 
perform the summation over all internal indices $ m \, , n \, \, \ldots \ne p,q $, then a 
simple power counting shows that the answer will be proportional to
\[
   \frac{\Omega^{m}}{\Omega^2} 
       \left( \frac{{\cal B}}{\Omega} \right)^{ 2 ( k_1 + k_2 + \ldots ) }
\]
(cf. Eqs.(\ref{G2-2Col-ser},\ref{G2-diag-2Col-ser}) and (\ref{G2-3Col-ser},\ref{G2-diag-3Col-ser})). 
Here $ \, \Omega^{m} \, $ in the numerator and $ \, \Omega^{2} \, $ in the denominator
result from the integration measure and from the factor $ \, {\cal R \, A} $, respectively. We
can rewrite this ratio as follows
\[
    {\cal B}^{ m - 2 }
       \left( \frac{{\cal B}}{\Omega} \right)^{ 2 ( k_1 + k_2 + \ldots ) - ( m - 2 )} .
\]
Obviously, having performed the summation over $ \, k_j $, the answer for $ \, {\cal G}^{(m)} 
\, $ can be  written as a product
\begin{equation}\label{FinVerif}
    {\cal G}^{(m)}( {\cal B}, \Omega ) = {\cal B}^{ m - 2 } \ 
                                                         \bar{{\cal G}}^{(m)} \left( \frac{\Omega}{{\cal B}} \right) \, ,
\end{equation}
which agrees with the estimate (\ref{VirCoefRat}) and with the formula (\ref{VE-coef-def}).

We remind that the $ \, m $-th virial coefficient $ \, \bar{{\cal G}}^{(m)} \, $ depends 
on the parameter $ \, \Omega/{\cal B} $, which can take on an arbitrary value and is not 
assumed  to be either small or large. The  successive terms of the VE decrease with 
increasing $ \, m \, $ only if the absolute value of the virial coefficients is bounded for the 
arbitrary ratio $ \, \Omega/ {\cal B} $.

\section{Conclusions}

In the present work we develop a supersymmetric field theoretical
description of a Gaussian ensemble of the almost diagonal Hermitian Random Matrices. 
In this ensemble the off-diagonal matrix elements are assumed to be parametrically smaller 
than the diagonal ones: $ H_{ii} \sim 1 \, , \ \, H_{ij} / H_{ii} \sim {\cal B} \ll  1 $. We  use
 the method of the supersymmetry to perform an ensemble averaging. The standard 
route of the derivation of the supersymmetric nonlinear $\sigma$-model can not
be taken in this case, since the diffusion approximation fails.

As an alternative to the supersymmetric $ \, \sigma$-model, we derive  a virial expansion 
(VE) in the number of ``interacting'' supermatrices, which is controlled by the small 
parameter $ \, {\cal B} \, $. Each supermatrix can be related to a localized eigenstate of 
the diagonal part of RMs. Thus, the supermatrix interaction describes the interaction of the 
localized wavefunctions via the (small) off-diagonal elements of RMs. The principle idea of VE 
is similar to one used in VE based on the Trotter  formula \cite{Trotter}. Nevertheless, the 
supersymmetric VE is much more powerful since it allows  to study not only the spectral 
correlations but also the correlation of wavefunctions taken at different energies and in 
different space points. 

The application of supersymmetric VE becomes especially efficient in a situation, when 
(i) the relevant energy range is much smaller than the typical value of the diagonal
elements of RMs, $ \, \omega \ll 1 $; and (ii)  the large scale approximation can be used, 
see Sect.\ref{SectVirCoef}. In this case the massive degrees of freedom are integrated out 
by the saddle-point approximation. This step is a counterpart of the saddle-point approximation 
used in the derivation  of the nonlinear $ \sigma $-model. However, the saddle-point 
approximation in the VE requires only  the linear constraint:
$ \, {\rm Str} [ Q ] = 0 $.

One of the main result of the paper is the integral expression for the $ \, m $-th term of 
VE,  Eqs.(\ref{Gm-L-scale},\ref{Gm-diag-L-scale}), which is governed by the interaction 
of $ \, m \, $ supermatrices. The superintegrals in this formula completely circumvent a 
complicated combinatorial calculations in the theory based on the Trotter formula. We 
note in passing that in this way we manage to reduce the complicated problem 
of simultaneous coloring of edges of several graphs (which is along standing problem
in  the statistical physics and the applied mathematics) to the calculation of superintegrals.
The superintegrals are calculated explicitly for the cases of 2- and 3-matrix interaction with 
the help of the parametrization suggested in Ref.[\onlinecite{SashaKr}]. The results 
containing in  Eqs.(\ref{G2-2Col-ser},\ref{G2-diag-2Col-ser}), 
(\ref{G2-2Col-final},\ref{G2-diag-2Col-final}) and (\ref{G2-3Col-ser},\ref{G2-diag-3Col-ser}) 
 have been obtained for the first time. They have been derived for a generic ensemble 
of the almost diagonal RMs described by Eq.(\ref{RMTdef}) in the case of the unitary symmetry 
class. We note that our approach can be easily generalized to the other symmetry classes.

The virial expansion generates a regular perturbation theory in powers of $ \, {\cal B } \, $ 
for a variety of the correlation functions in the different models of the almost diagonal RMs. 
Applications to certain RMT models will be presented elsewhere. The critical ensemble of 
RMs \cite{Chi-Virial} and the Moshe-Neuberger-Shapiro model \cite{MNS} with the orthogonal 
symmetry are two examples of promising applications. The next important step would be derivation 
of non-perturbative results directly from the supersymmetric action in the large scale 
approximation. The non-perturbative solutions could particularly shed light on the following
problem: under what circumstances an interaction between the localized states can lead 
to the criticality or to the delocalization. This question is of fundamental importance in the
 theory of disordered \cite{Ziman} and strongly-correlated disordered systems \cite{BAA}.

\acknowledgments

The authors are very grateful to Vladimir Kravtsov for initiating the project
on the field theory representation of almost diagonal RMs and critical
comments concerning the accuracy of VE and to Vladimir
Yudson for useful discussions. A.O. acknowledges support by the 
Dutch Science Foundation NWO/FOM.

\appendix

\section{Parametrization of matrix $ \, Q $ in the retarded-advanced sector} \label{Q-param}

Let us find a parametrization of the matrix $Q$
defined as direct product of the supervector  by the conjugated supervector
\begin{eqnarray}\label{Q-def}
Q\equiv\Psi\otimes\bar{\Psi}= \left(\begin{matrix} \Psi_R\otimes\Psi_R^\dagger &
\Psi_R\otimes K\Psi_A^\dagger \\ \Psi_A\otimes\Psi_R^\dagger &\Psi_A\otimes K\Psi_A^\dagger 
\end{matrix}\right), \\ 
\Psi = \left(\begin{matrix} \Psi_R\\ \Psi_A
\end{matrix}\right),
\ 
\Psi_{R/A} = \left(\begin{matrix} s_{R/A}\\ \chi_{R/A}
\end{matrix}\right),
\ 
K\equiv \left(\begin{matrix} -1 & 0\\0 & 1
 \end{matrix}\right),
\end{eqnarray}
where indices $R$ and $A$ are referring to the retarded and advanced
sectors correspondingly.
 
Matrix $Q_{RR}= \Psi_R\otimes\Psi_R^\dagger$ is the orthogonal projector on
vector $\Psi_R$ and can be diagonalized by the unitary matrix $U_R$:
\begin{eqnarray}
\label{u1}
&Q_{RR}=U_RD_{RR}U_R^{-1} & \\
&U_R=\left(\begin{matrix} 1-\frac{1}{2}\eta_R^\ast\eta_R & -\eta_R^\ast\\
\eta_R &  1+\frac{1}{2}\eta_R^\ast\eta_R \end{matrix}\right),
\  U^{-1}_R=\left(\begin{matrix} 1-\frac{1}{2}\eta_R^\ast\eta_R & \eta_R^\ast\\
-\eta_R &  1+\frac{1}{2}\eta_R^\ast\eta_R \end{matrix}\right),
\  D_{RR}=\left(\begin{matrix} \lambda_R^2 & 0\\
0 & 0 \end{matrix}\right)&
\nonumber
\end{eqnarray}
where $\eta_R=\chi_R/s_R$, $\lambda_R^2=|| \Psi_R ||^2=|s_R|^2 +\chi_R^\ast 
\chi_R$. In the similar way block $Q_{AA}$ can be diagonalized by the pseudounitary
matrix $U_A$ ($U_A^\dagger KU_A=K$):
\begin{eqnarray}
\label{u2}
&Q_{AA}=U_AD_{AA}U_A^{-1} & \\
&U_A=\left(\begin{matrix} 1+\frac{1}{2}\eta_A^\ast\eta_A & \eta_A^\ast\\
\eta_A &  1-\frac{1}{2}\eta_A^\ast\eta_A \end{matrix}\right),
\ U^{-1}_A=\left(\begin{matrix} 1+\frac{1}{2}\eta_A^\ast\eta_A & -\eta_A^\ast\\
-\eta_A &  1-\frac{1}{2}\eta_A^\ast\eta_A \end{matrix}\right),
\ D_{AA}=\left(\begin{matrix} -\lambda_A^2 & 0\\
0 & 0 \end{matrix}\right)&
\nonumber
\end{eqnarray}
with  $\eta_A=\chi_A/s_A$, $\lambda_A^2=- || \Psi_A ||^2=|s_A|^2
-\chi_A^\ast \chi_A$. Moreover $U_R, U_A$ diagonalize off-diagonal blocks of $Q$:
\begin{eqnarray}
\label{offdiag}
&Q_{AR}=U_AD_{AR}U_R^{-1} \hspace{1cm} Q_{RA}=U_RD_{RA}U_A^{-1} &\nonumber\\
& D_{AR}=\left(\begin{matrix} e^{i\phi}\lambda_R\lambda_A & 0\\
0 & 0 \end{matrix}\right)\hspace{1cm} D_{RA}=\left(\begin{matrix} -e^{-i\phi}\lambda_R\lambda_A & 0\\
0 & 0 \end{matrix}\right)&,
\end{eqnarray}
where $e^{i\phi}=s_R^\ast s_A/(\mid s_R s_A\mid)$. Thus the matrix $ \, Q \, $ can be
parameterized as follows:
\begin{eqnarray}
\label{Param}
&Q=UDU^{-1} &\nonumber\\
&U=\left(\begin{matrix}U_R & 0\\
0 &  U_{A}\end{matrix}\right),
\hspace{1cm} D=\left(\begin{matrix} D_{RR} & D_{RA}\\
D_{AR} & D_{AA}\end{matrix}\right)=\left(\begin{matrix} 
\lambda_R^2 & 0 &  -e^{-i\phi}\lambda_R\lambda_A  & 0\\
0 & 0 & 0 & 0 \\
e^{i\phi}\lambda_R\lambda_A  & 0 & -\lambda_A^2 & 0 \\
0 & 0 & 0 & 0 \end{matrix}\right)&.
\end{eqnarray}
The measure in this parametrization can be easily found by calculating the Jacobian
(Berezenian) of the transformation (\ref{Param}) and is equal to
\begin{equation}\label{NewMeasure}
 {\cal D} \, \{ Q \} =  \frac{2}{\pi} \frac{d\lambda_R d\lambda_A}{\lambda_R \lambda_A}
                                    d\phi d\eta_R^\ast d\eta_R d\eta_A^\ast d\eta_A \, .
\end{equation} 
Using Eq.(\ref{Param}) and taking into account the rotational symmetry of the supertrace, 
we obtain:
\begin{eqnarray}
\label{Str-Terms-1}
   {\rm Str} (Q)  &=&  R \, , \  
   {\rm Str} (Q^2) = R^2 \, ; \quad R \equiv \lambda_R^2 - \lambda_A^2 \, ; \\
   {\rm Str} (\Lambda Q) &=& S \, ; \quad S \equiv \lambda_R^2 + \lambda_A^2 \, ;
\label{Str-Terms-2}
\end{eqnarray}
\begin{eqnarray}\label{Str-Prod}
   {\rm Str} [ Q\tilde{Q} ] & = &
       \lambda_R^2\tilde{\lambda}_{R}^{2}
               (1-\alpha_R^\ast\alpha_R) +
       \lambda_A^2\tilde{\lambda}_A^{2}
               (1+\alpha_A^\ast\alpha_A) \\
     & & 
     -2\cos \theta \lambda_R\tilde{\lambda}_R\lambda_A\tilde{\lambda}_A
           \left(1-\frac{1}{2}\alpha_R^\ast\alpha_R \right)
           \left(1+\frac{1}{2}\alpha_A^\ast\alpha_A \right) \, ;
\nonumber
\end{eqnarray} 
where $ \, \alpha_{R/A}\equiv\eta_{R/A}-\tilde{\eta}_{R/A} \, $; $ \, \theta \equiv \phi-\tilde{\phi}+
\Delta \, $; $\Delta=\frac{i}{2}(\tilde{\eta}_R^{\ast}\eta_R- \eta_R^{\ast}\tilde{\eta}_R
+ \tilde{\eta}_A^{\ast}\eta_A-\eta_A^{\ast}\tilde{\eta}_A)$ and tilde marks the
variables of the matrix $ \, \tilde{Q} $.  The last identity (\ref{Str-Prod}) follows from the well known 
property \cite{Zirnbauer86} of the matrices $ \, U $ :
\begin{eqnarray}
U_R^{-1}(\tilde{\eta}_R)U_R(\eta_R)&=&U_R(\eta_R-\tilde{\eta}_R)
         e^{\frac{1}{2}(\tilde{\eta}_R^{\ast}\eta_R-
         \eta_R^{\ast}\tilde{\eta}_R)}  , \\
U_A^{-1}(\tilde{\eta}_A)U_A(\eta_A)&=&U_A(\eta_A-\tilde{\eta}_A)
         e^{-\frac{1}{2}(\tilde{\eta}_A^{\ast}\eta_A-
         \eta_A^{\ast}\tilde{\eta}_A)} .
\end{eqnarray}

\section{Parametrization of matrix $ \, Q $ in the retarded-retarded sector} \label{QRR-param}

In the retarded-retarded sector the matrix $Q$ is defined similar to Eq.(\ref{Q-def}) but without
matrix $K$:
\begin{eqnarray}
Q\equiv\Psi\otimes\Psi^\dagger= \left(\begin{matrix} \Psi_R\otimes\Psi_{R'}^\dagger &
\Psi_R\otimes \Psi_{R'}^\dagger \\ \Psi_{R'}\otimes\Psi_R^\dagger &\Psi_{R'}\otimes \Psi_{R'}^\dagger 
\end{matrix}\right), \\ 
\Psi = \left(\begin{matrix} \Psi_R\\ \Psi_{R'}
\end{matrix}\right),
\ 
\Psi_{R/R'} = \left(\begin{matrix} s_{R/R'}\\ \chi_{R/R'}
\end{matrix}\right),
\end{eqnarray}
where indices $R$ and $R'$ are referring to  the  retarded
sectors of two different Green's functions. As a result one can diagonalize
$Q$ by transformation similar to Eq.(\ref{Param}):
\begin{eqnarray}\label{QRR-par}
&Q=UDU^{-1} &\nonumber\\
&U=\left(\begin{matrix}U_R & 0\\
0 &  U_{R'}\end{matrix}\right),
\hspace{1cm} D=\left(\begin{matrix} D_{RR} & D_{R{R'}}\\
D_{{R'}R} & D_{{R'}{R'}}\end{matrix}\right)=\left(\begin{matrix} 
\lambda_R^2 & 0 &  e^{-i\phi}\lambda_R\lambda_{R'}  & 0\\
0 & 0 & 0 & 0 \\
e^{i\phi}\lambda_R\lambda_{R'}  & 0 & \lambda_{R'}^2 & 0 \\
0 & 0 & 0 & 0 \end{matrix}\right)&,
\end{eqnarray}
where $U_R$ is defined in Eq.(\ref{u1}) and $U_{R'}$ is obtained from
$U_R$ by replacing subscript $R$ by $R'$ everywhere. All parameters appearing
in (\ref{QRR-par}) are defined in the same way as before:
\begin{eqnarray}
 \eta_R&=&\frac{\chi_R}{s_R},\;\; \lambda_R^2=|s_R|^2 +\chi_R^\ast \chi_R,\;\;
e^{i\phi}=\frac{s_R^\ast s_{R'}}{\mid s_R s_{R'}\mid},\nonumber\\
\eta_{R'}&=&\frac{\chi_{R'}}{s_{R'}},\;\; \lambda_{R'}^2=|s_{R'}|^2 +\chi_{R'}^\ast \chi_{R'}
\end{eqnarray} 
 
The measure in this parametrization remains the same as in Eq.(\ref{NewMeasure}), while
the roles of variables $R$ and $S$ defined in Eq.(\ref{Str-Terms-1}) and Eq.(\ref{Str-Terms-2})
are interchanged now:
\begin{eqnarray}
\label{Str-Terms-1-RR}
   {\rm Str} (Q)  &=&  S \, , \  
   {\rm Str} (Q^2) = S^2 \, ; \quad S \equiv \lambda_R^2 + \lambda_{R'}^2 \, ; \\
   {\rm Str} (\Lambda Q) &=& R \, ; \quad R \equiv \lambda_R^2 - \lambda_{R'}^2 \, .
\label{Str-Terms-2-RR}
\end{eqnarray}
Finally the expression for ${\rm Str} [ Q\tilde{Q} ]$ is again similar to Eq.(\ref{Str-Prod}):
\begin{eqnarray}\label{Str-Prod-RR}
   {\rm Str} [ Q\tilde{Q} ] & = &
       \lambda_R^2\tilde{\lambda}_{R}^{2}
               (1-\alpha_R^\ast\alpha_R) +
       \lambda_{R'}^2\tilde{\lambda}_{R'}^{2}
               (1-\alpha_{R'}^\ast\alpha_{R'}) \\
     & & 
     -2\cos \theta \lambda_R\tilde{\lambda}_R\lambda_{R'}\tilde{\lambda}_{R'}
           \left(1-\frac{1}{2}\alpha_R^\ast\alpha_R \right)
           \left(1-\frac{1}{2}\alpha_{R'}^\ast\alpha_{R'} \right) \, ;
\nonumber
\end{eqnarray} 
where $ \, \alpha_{R/{R'}}\equiv\eta_{R/{R'}}-\tilde{\eta}_{R/{R'}} \, $; $ \, \theta \equiv \phi-\tilde{\phi}+
\Delta \, $; $\Delta=\frac{i}{2}(\tilde{\eta}_R^{\ast}\eta_R- \eta_R^{\ast}\tilde{\eta}_R
- \tilde{\eta}_{R'}^{\ast}\eta_{R'}+\eta_{R'}^{\ast}\tilde{\eta}_{R'})$.

\section{Integrals for integer powers of $ \, \sin(\phi/2) $}
\label{PhaseIntCalc}

One needs the following formulas to average a product of the 
supertraces over the phases:
\begin{eqnarray}\label{Ph-Int-1}
   {\cal F}(k) & \equiv &
   \frac{1}{2\pi} \int_0^{2\pi} \sin^{2k} \left( \frac{\phi}{2} \right) {\rm d} \phi =
         \frac{\Gamma(k+1/2)}{\pi^{1/2} \, \Gamma(k+1)} \, ; \\
\label{Ph-Int-2}
   {\cal F}(k_1,k_2,k_3) & \equiv &
   \frac{1}{(2\pi)^2} \int \!\!\! \int_0^{2\pi} 
        \sin^{2k_1} \left( \frac{\phi_1}{2} \right) 
        \sin^{2k_2} \left( \frac{\phi_2}{2} \right) 
        \sin^{2k_3} \left( \frac{\phi_1-\phi_2}{2} \right) 
   {\rm d} \phi_{1,2} = \\
              \frac{\Gamma(k_1 + 1/2)}{\pi^{1/2}} 
     & & \!\!\!\!\!\!\!\!\!
              \frac{\Gamma(k_2 + 1/2)}{\pi^{1/2}}
              \frac{\Gamma(k_3 + 1/2)}{\pi^{1/2}}
                \frac{\Gamma(k_1 + k_2 +k_3 +1)}
        {\Gamma(k_1+k_2 +1) \Gamma(k_1+k_3 +1) \Gamma(k_2+k_3 +1)} \, .
\nonumber
\end{eqnarray}
We consider only integer powers of sines $ \, k \, $ and $ \, k_{1,2,3} $.
Eq.(\ref{Ph-Int-1}) can be found in standard mathematical tables \cite{R-Gr} while
Eq.(\ref{Ph-Int-2}) can be proven by the induction over one of the exponents, for 
example, over $ \, k_3 $. Clearly, $ \, {\cal F}(k,0,0) = {\cal F}(k) \, ; \ {\cal F}(k_1,k_2,0) 
= {\cal F}(k_1) {\cal F}(k_2) $.  This constitutes the induction basis. 
To check the hypothesis for arbitrary $ \, 
k_3 $, we assume that $ \, {\cal F} (k_1',k_2', k_3) \, $ is known
for arbitrary $ \, k_{1,2}' \, $ and find a relation between $ \, {\cal F}(k_1,k_2, k_3+1) \, $ 
and $ \, {\cal F}(k_1',k_2', k_3) $. Simple trigonometric
transformations together with integrations by parts  yield
\begin{eqnarray}\label{Finduc}
   {\cal F}(k_1,k_2, k_3+1) & = & {\cal F}(k_1+1,k_2, k_3) + {\cal F}(k_1,k_2+1, k_3)
                   - 2 {\cal F}(k_1+1,k_2+1, k_3) + \\
        & &
                   + \frac{2}{(k_1+1)(k_2+1)} {\cal F}^{(2)}(k_1+1,k_2+1,k_3) \, ,
\nonumber
\end{eqnarray}
where
\begin{eqnarray}\label{Ph-Int-3}
   {\cal F}^{(p)}(k_1,k_2,k_3) & \equiv &
   \frac{1}{(2\pi)^2} \int \!\!\! \int_0^{2\pi} 
    \left\{
    \partial^{p}_{\phi_1}
     \left[
        \sin^{2k_1} \left( \frac{\phi_1}{2} \right) 
    \right]
   \right\}
        \sin^{2k_2} \left( \frac{\phi_2}{2} \right) 
        \sin^{2k_3} \left( \frac{\phi_1-\phi_2}{2} \right) \!\! , \\
   {\cal F}^{(2)}(k_1,k_2,k_3) & = &
      \frac{k_1}{2} \biggl[ 
             (2 k_1 - 1) {\cal F}(k_1-1,k_2, k_3) - 2 k_1 {\cal F}(k_1,k_2, k_3)
                               \biggr] \, .
\label{F2}
\end{eqnarray}
Inserting (\ref{F2}) into (\ref{Finduc}), we get the relation
\begin{eqnarray}\label{InducRel}
   {\cal F}(k_1,k_2, k_3+1) & = & {\cal F} (k_1+1,k_2,k_3) +
                    \left( 1 + \frac{2k_1+1}{k_2+1} \right) {\cal F} (k_1,k_2+1,k_3) - \\
     & & \ 
                  -2 \left( 1 + \frac{k_1+1}{k_2+1}  \right) {\cal F} (k_1+1,k_2+1,k_3) \, .
\nonumber
\end{eqnarray}
We substitute Eq.(\ref{Ph-Int-2}) in the right-hand side of  Eq.(\ref{InducRel})
and derive the answer
\begin{eqnarray}
   {\cal F}(k_1,k_2,k_3+1) & = & 
              \frac{\Gamma(k_1 + 1/2)}{\pi^{1/2}}
              \frac{\Gamma(k_2 + 1/2)}{\pi^{1/2}}
              \frac{\Gamma(k_3 + 3/2)}{\pi^{1/2}}  \times \\
      & & \quad
                \frac{\Gamma(k_1 + k_2 +k_3 +2)}
        {\Gamma(k_1+k_2 +1) \Gamma(k_1+k_3 +2) \Gamma(k_2+k_3 +2)} \, ,
\nonumber
\end{eqnarray}
which satisfies Eq.(\ref{Ph-Int-2}). Thus the
induction over $ \, k_3 \, $ is verified and Eq.(\ref{Ph-Int-2}) is proven.

Using (\ref{Ph-Int-2},\ref{Ph-Int-3},\ref{F2}), one can also show that
\begin{eqnarray}
\label{Ph-Int-4}
   {\cal F}^{(2n - 1)}(k_1,k_2,k_3) & = & 0, \quad n = 1, 2, \ldots \\
\label{Ph-Int-5}
   {\cal F}^{(2)}(k_1,k_2,k_3) & = & {\cal F}(k_1,k_2,k_3) \frac{k_1 k_2 k_3}{k_1 + k_2 + k_3} \, ,
\end{eqnarray}
and
\begin{eqnarray}
\label{Ph-Int-6}
   {\cal F}^{(4)}(k_1,k_2,k_3) & = &
       \frac{k_1}{8} \Bigl[
         2(k_1-1)(2 k_1 - 1)(2 k_1 - 3) {\cal F}(k_1-2,k_2,k_3) - \cr
     & & \ 
        - 4 ( 4 k_1^3 - 6 k_1^2 + 4 k_1 - 1 ) {\cal F}(k_1-1,k_2,k_3) +
          8 k_1^3 {\cal F}(k_1,k_2,k_3)
                                \Bigr] = \cr
     & = & 
      - {\cal F}^{(2)}(k_1,k_2,k_3) 
                                  \left(
                         1 - \frac{ ( k_1 - 1 ) ( k_2 - 1 ) ( k_3 - 1) }{ k_1 + k_2 + k_3 - 1 }
                                  \right) \, .
\end{eqnarray}

\section{Averaging $ \, {\rm Str}^k [Q_1 Q_2] \, $  and 
               $ \, {\rm Str}^{k_1} [Q_1 Q_2] {\rm Str}^{k_2} [Q_1 Q_3] {\rm Str}^{k_3} [Q_2 Q_3]\, $ 
              over the phases at $ \, R_{1,2,3} = 0 $ and integers powers $ \, k, \, k_{1,2,3} $}
\label{PhAver}

Let us average $ \, {\cal S}^k_{12} \equiv \left( {\rm Str} [Q_1 Q_2] \right)^k \Bigl|_{R_{1,2}=0} $ 
over the phases $ \, \phi_{1,2} $:
\begin{equation}\label{Av-1}
    {\cal S}^k_{12} = 
       \left[
        \frac{S_1 S_2}{4}
          \left(
            4 \sin^2\left( \frac{\phi}{2} \right) + 
            2 (\alpha_A^*\alpha_A - \alpha_R^*\alpha_R) \sin^2\left( \frac{\phi}{2} \right)  +
            \frac{\cos(\phi)}{2} \alpha_A^*\alpha_A \alpha_R^*\alpha_R
          \right)
       \right]^k \, ;
\end{equation}
where $ \, k \, $ is integer; $ \, \phi \equiv \phi_{12} + \Delta_{12} \, ; \ \phi_{12} \equiv \phi_1 - \phi_2 
\, $ and indices $ \, 1 \, $ and $ \, 2 \, $ mark the variables of the matrices $ \, Q_1 \, $ and $ \, Q_2 $, 
respectively (see also notations in the Appendix \ref{Q-param}). We have to calculate 
the following integral of the periodic function:
\begin{equation}\label{OnePhAver}
   \frac{1}{(2\pi)^2}
   \int \!\!\! \int_0^{2\pi} {\rm d} \phi_{1,2} \, {\cal S}^k_{12} ( \phi_1-\phi_2+ \Delta_{12})=
      \frac{1}{2\pi} \int_0^{2\pi} {\rm d} \phi_{12} \, {\cal S}^k_{12} (\phi_{12}) \, .
\end{equation}
The nilpotents from $ \, \Delta_{12} \equiv \, \frac{\imath}{2} ( \Delta^{(12)}_R 
+ \Delta^{(12)}_A ) $, $ \, \Delta^{(12)}_{R/A} = ( \eta_{R/A} )^*_2 \, ( \eta_{R/A} )_1 
- c.c. $, give no contribution to the integral (\ref{OnePhAver}) due to the periodicity of 
the integrand.

Collecting the terms with the same powers of the Grassmann variables we find
\begin{eqnarray}\label{Av-2}
    {\cal S}^k_{12} & = & \left( \frac{S_1 S_2}{4} \right)^k
       \left\{
              \Bigl( 2 \sin \left( \phi/2 \right) \Bigr)^{2k} +
              \frac{k}{2} \Bigl( 2 \sin \left( \phi/2 \right) \Bigr)^{2k}
              ( \alpha_A^*\alpha_A - \alpha_R^*\alpha_R ) + 
      \right. \\
       & & \quad + \left. 
                 \frac{k}{2}   \left[  
                                   \Bigl( 2 \sin \left( \phi/2 \right) \Bigr)^{2(k-1)} 
         - \frac{k}{2} \Bigl( 2 \sin \left( \phi/2 \right) \Bigr)^{2k}
                                        \right] 
               \alpha_A^* \alpha_A \alpha_R^* \alpha_R 
        \right\} \, .
\nonumber
\end{eqnarray}
We insert Eq.(\ref{Av-2}) into the integral  (\ref{OnePhAver}) and integrate
over the phase $ \, \phi_{12} \, $ using Eq.(\ref{Ph-Int-1}):
\begin{eqnarray}\label{Av-3}
   \frac{1}{2\pi}
   \int_0^{2 \pi} \!\!\! {\rm d} \phi_{12} \, 
      {\cal S}^k_{12}( \phi_{12} )
                   & = & \left ( \frac{ S_1 S_2 }{4} \right)^k
                               \frac{\Gamma(2k-1)}{\Gamma(k-1)\Gamma(k)} \times \\
         & \times & \Biggl\{
        \frac{2k-1}{k(k-1)}
           \Bigl( 2 + k (\alpha_A^*\alpha_A - \alpha_R^*\alpha_R) \Bigr) - 
           k \alpha_A^*\alpha_A \alpha_R^*\alpha_R 
                            \Biggr\} \, .
\nonumber
\end{eqnarray}
Note that the equality (\ref{Av-2}) can be rewritten in a more compact form:
\begin{equation}\label{Av-4}
    {\cal S}^k_{12} = \left( S_1 S_2 \right)^k
      \left\{ 
         \Bigl( \sin \left( \phi/2 \right) \Bigr)^{2k} e^{ k \Upsilon_{12} } - 
         \frac{k}{4} \Bigl( \sin \left( \phi/2 \right) \Bigr)^{2(k-1)} \Upsilon_{12}^2 
      \right\} \, ; 
    \Upsilon_{12} \equiv  \frac{\alpha_A^*\alpha_A -  \alpha_R^*\alpha_R}{2} \ .
\end{equation}

Unlike Eq.(\ref{OnePhAver}) written for the case of 2 linked supermatrices, the averaged product 
$ \, {\cal S}^{k_1}_{12} \, {\cal S}^{k_2}_{13} \, {\cal S}^{k_3}_{23} \, $ (which includes
3 linked supermatrices) depends on the nilpotents coming from $ \, \Delta^{(3)} \equiv \, 
\Delta_{12} + \Delta_{13} + \Delta_{23} \, $ and it can be expanded in the even powers
of $ \, \Delta^{(3)} $:
\begin{eqnarray}
\label{2PhAver-1}
   \frac{1}{(2\pi)^3} \int \!\!\!\! \int \!\!\!\! \int_0^{2\pi} \!\!\!\! {\rm d} \phi_{1,2,3} \, 
         {\cal S}^{k_1}_{12} \left( \phi_1 - \phi_2 + \Delta_{12} \right)
    & & \!\!\!\!\!
         {\cal S}^{k_2}_{13} \left( \phi_1 - \phi_3 + \Delta_{13} \right) 
         {\cal S}^{k_3}_{23} \left( \phi_2 - \phi_3 + \Delta_{23} \right) = \\
   \frac{1}{(2\pi)^2} \int \!\!\!\! \int_0^{2\pi} \!\!\!\! {\rm d} \phi \, {\rm d} \phi' \, 
         {\cal S}^{k_1}_{12} \left( \phi \right)
    & & \!\!\!\!\!
         {\cal S}^{k_2}_{13} \left( \phi' \right) 
         {\cal S}^{k_3}_{23} \left( \phi - \phi' + \Delta^{(3)} \right) = \cr
\label{2PhAver-2}
   \sum_{p=0}^3 \frac{ \left( \Delta^{(3)} \right)^{2p} }{ (2p)! }
 \Biggl[ 
   \frac{1}{(2\pi)^2} \int \!\!\!\! \int_0^{2\pi} \!\!\!\! {\rm d} \phi \, {\rm d} \phi' \, 
         {\cal S}^{k_1}_{12} \left( \phi \right)
    & & \!\!\!\!\! 
         {\cal S}^{k_2}_{13} \left( \phi' \right) 
         \partial^{2p}_{\phi} {\cal S}^{k_3}_{23} \left( \phi - \phi'  \right) 
 \Biggr] \, .
\end{eqnarray}
Here $ \, k_{1,2,3} \, $ are integer.
There are no odd powers of $ \, \Delta^{(3)} \, $ due to the equality (\ref{Ph-Int-4}). For
the purpose of the present paper, we need only the terms with $ \, p = 0, 1, 2 $. The term
$ \, \propto \left( \Delta^{(3)} \right)^{6} \, $ yields zero after multiplication by the factor
$ \, {\cal R}_p \, {\cal A}_q $ (see Eqs.(\ref{Gm-diag-L-scale},\ref{Gm-L-scale})).
Using the formula (\ref{Av-4}) and the results of Appendix \ref{PhaseIntCalc}, we find:

\begin{eqnarray}\label{2PhAver-3}
\frac{ \left( \Delta^{(3)} \right)^{2p} }{ (2p)! }
    & & \!\!\!\!\! 
 \Biggl[ 
   \frac{1}{(2\pi)^2} \int \!\!\!\! \int_0^{2\pi} \!\!\!\! {\rm d} \phi \, {\rm d} \phi' \,
         {\cal S}^{k_1}_{12} \left( \phi \right)
         {\cal S}^{k_2}_{13} \left( \phi' \right) 
         \partial^{2p}_{\phi} {\cal S}^{k_3}_{23} \left( \phi - \phi'  \right) 
 \Biggr] \Biggl|_{p=2} = \\
  & & =
    ( S_1 S_2 )^{k_1}  ( S_1 S_3 )^{k_2} ( S_2 S_3 )^{k_3}
      \frac{ \left( \Delta^{(3)} \right)^4 }{4!} \, {\cal F}^{(4)}( k_1, k_2, k_3 ) \, ;
\nonumber
\end{eqnarray}

\begin{eqnarray}\label{2PhAver-4}
    & & \!\!\!\!\! 
\frac{ \left( \Delta^{(3)} \right)^{2p} }{ (2p)! }
 \Biggl[ 
   \frac{1}{(2\pi)^2} \int \!\!\!\! \int_0^{2\pi} \!\!\!\! {\rm d} \phi \, {\rm d} \phi' \, 
         {\cal S}^{k_1}_{12} \left( \phi \right)
         {\cal S}^{k_2}_{13} \left( \phi' \right) 
         \partial^{2p}_{\phi} {\cal S}^{k_3}_{23} \left( \phi - \phi'  \right) 
 \Biggr] \Biggl|_{p=1} = \\
  & & =
    ( S_1 S_2 )^{k_1}  ( S_1 S_3 )^{k_2} ( S_2 S_3 )^{k_3}
      \frac{ \left( \Delta^{(3)} \right)^2 }{8}  \Bigl(
        2 {\cal F}^{(2)}( k_1, k_2, k_3 ) \left( k_1 \Upsilon_{12} + k_2 \Upsilon_{13} + 
                                                                          k_3 \Upsilon_{23} \right)^2 - \cr
  & & - \left [ 
      {\cal F}^{(2)}( k_1-1, k_2, k_3 ) k_1 \Upsilon_{12}^2 +
      {\cal F}^{(2)}( k_1, k_2-1, k_3 ) k_2 \Upsilon_{13}^2 +
      {\cal F}^{(2)}( k_1, k_2, k_3 -1) k_3 \Upsilon_{23}^2
            \right]
  \Bigr) \, ;
\nonumber
\end{eqnarray}

\begin{eqnarray}\label{2PhAver-5}
\frac{ \left( \Delta^{(3)} \right)^{2p} }{ (2p)! }
    & & \!\!\!\!\! 
 \Biggl[ 
   \frac{1}{(2\pi)^2} \int \!\!\!\! \int_0^{2\pi} \!\!\!\! {\rm d} \phi \, {\rm d} \phi' \, 
         {\cal S}^{k_1}_{12} \left( \phi \right)
         {\cal S}^{k_2}_{13} \left( \phi' \right) 
         \partial^{2p}_{\phi} {\cal S}^{k_3}_{23} \left( \phi - \phi'  \right) 
 \Biggr] \Biggl|_{p=0} = \\
  & & =
    ( S_1 S_2 )^{k_1}  ( S_1 S_3 )^{k_2} ( S_2 S_3 )^{k_3}
  \bigl( Part_1 + Part_2 + Part_3 \bigr) \, ;
\nonumber
\end{eqnarray}
where
\begin{eqnarray}\label{2PhAver-6}
   Part_1 & = & 
    \frac{{\cal F}( k_1, k_2, k_3 )}{4} \Bigl( 
         2 \ k_1 k_2 k_3 \ \Upsilon_{1,2} \Upsilon_{1,3} \Upsilon_{2,3} \, 
                              [ 
                                k_1 \Upsilon_{1,2} +
                                k_2 \Upsilon_{1,3} +
                                k_3 \Upsilon_{2,3}
                              ] + \cr
  & & \quad + 
                                ( k_1 \Upsilon_{1,2} k_2 \Upsilon_{1,3} )^2 +
                                ( k_1 \Upsilon_{1,2} k_3 \Upsilon_{2,3} )^2 +
                                ( k_2 \Upsilon_{1,3} k_3 \Upsilon_{2,3} )^2
                                               \Bigr) \, ,
\nonumber
\end{eqnarray}
\begin{eqnarray}\label{2PhAver-7}
   Part_2 & = & - \frac{1}{8} \Bigl( 
      {\cal F}( k_1-1, k_2, k_3 ) k_1 \Upsilon_{1,2}^2 ( k_2 \Upsilon_{1,3} + k_3 \Upsilon_{2,3} )^2 + \\
   +
      {\cal F}( k_1, k_2-1, k_3 ) \!\! & k_2 & \!\! \Upsilon_{1,3}^2 ( k_1 \Upsilon_{1,2} + k_3 \Upsilon_{2,3} )^2 +
      {\cal F}( k_1, k_2, k_3-1 ) k_3 \Upsilon_{2,3}^2 ( k_1 \Upsilon_{1,2} + k_2 \Upsilon_{1,3} )^2
                                                   \Bigr) \, ,
\nonumber
\end{eqnarray}
\begin{eqnarray}\label{2PhAver-8}
   Part_3 & = &  \frac{1}{16} \Bigl( 
      {\cal F}( k_1-1, k_2-1, k_3 ) k_1 \Upsilon_{1,2}^2 k_2 \Upsilon_{1,3}^2 + \\
               & & +
      {\cal F}( k_1-1, k_2, k_3-1 ) k_1 \Upsilon_{1,2}^2 k_3 \Upsilon_{2,3}^2 +
      {\cal F}( k_1, k_2-1, k_3-1 ) k_2 \Upsilon_{1,3}^2 k_3 \Upsilon_{2,3}^2
                                                 \Bigr) \, .
\nonumber
\end{eqnarray}

\section{Results of the integration \protect\\
                over the Grassmann variables}\label{ThanksMaple} 

A direct integration over the large number of the Grassmann variables
(12 Grassmanns in the case of 3-matrix approximation) is technically trivial but very
long and boring arithmetic procedure. We have used the ``Grassmann'' package of
the ``Maple'' system to do this step of the calculations. Here we present the results 
of this procedure which are necessary for the calculations in the 3-matrix approximation
\begin{equation}
   \int {\rm d} \{ \eta^*_R \eta_R \eta^*_A \eta_A \}_{p,q,m} \,
      \bigl( \eta^*_R \eta_R \bigr)_p \bigl( \eta^*_A \eta_A \bigr)_q \, ( \Delta^{(3)} )^4 = 
      \frac{ 4! }{ 2^4 } \, ;
\end{equation}
\begin{equation}
   \int {\rm d} \{ \eta^*_R \eta_R \eta^*_A \eta_A \}_{p,q,m} \,
      \bigl( \eta^*_R \eta_R \bigr)_p \bigl( \eta^*_A \eta_A \bigr)_q \, ( \Delta^{(3)} )^2 \times
            \left\{
              \begin{array}{c}
                    \Upsilon^2_{p,q} \cr
                    \Upsilon^2_{p,m} \cr
                    \Upsilon^2_{q,m}
              \end{array}
            \right\} = 0 \, ; 
\end{equation}
\begin{equation}
   \int {\rm d} \{ \eta^*_R \eta_R \eta^*_A \eta_A \}_{p,q,m} \,
      \bigl( \eta^*_R \eta_R \bigr)_p \bigl( \eta^*_A \eta_A \bigr)_q \, ( \Delta^{(3)} )^2 \times
            \left\{
              \begin{array}{c}
                    \Upsilon_{p,q}  \Upsilon_{q,m} \cr
                    \Upsilon_{p,m} \Upsilon_{m,q} \cr
                    \Upsilon_{q,p} \Upsilon_{p,m}
              \end{array}
            \right\} = \frac{1}{2^2} \, ; 
\end{equation}
\begin{equation}
   \int {\rm d} \{ \eta^*_R \eta_R \eta^*_A \eta_A \}_{p,q,m} \,
      \bigl( \eta^*_R \eta_R \bigr)_p \bigl( \eta^*_A \eta_A \bigr)_q \times
            \left\{
              \begin{array}{c}
                    \Upsilon^2_{p,q}  \Upsilon^2_{q,m} \cr
                    \Upsilon^2_{p,m} \Upsilon^2_{m,q} \cr
                    \Upsilon^2_{q,p} \Upsilon^2_{p,m}
              \end{array}
            \right\} = \frac{1}{2^2} \, ; 
\end{equation}
\begin{equation}
   \int {\rm d} \{ \eta^*_R \eta_R \eta^*_A \eta_A \}_{p,q,m} \,
      \bigl( \eta^*_R \eta_R \bigr)_p \bigl( \eta^*_A \eta_A \bigr)_q \times
            \left\{
              \begin{array}{c}
                    \Upsilon^2_{p,q}  \Upsilon_{q,m} \Upsilon_{m,p} \cr
                    \Upsilon_{p,q}  \Upsilon^2_{q,m} \Upsilon_{m,p} \cr
                    \Upsilon_{p,q}  \Upsilon_{q,m} \Upsilon^2_{m,p} 
              \end{array}
            \right\} = \frac{1}{2^2} \, ; 
\end{equation}
see the definitions of the nilpotents in the previous Appendix. Note that all 3 indices
$ \, p, q \, $ and $ \, m \, $ are different.


\begin{thebibliography}{50}
\bibitem{Psi-Q} Note that we have transformed the action written in terms of
                     the supervectors $ \, \Psi \, $ with 4 commuting and 4 anticommuting 
                     variable to the action written in terms of the supermatrices $ \, Q \, $
                     with 3 commuting and 4 anticommuting variable. One phase of the
                     commuting variables has been integrated out since the direct product 
                     (\ref{Q-blocks}) does not depend on it. This is possible for RMTs with 
                     uncorrelated matrix elements having zero mean value, cf. Eq.(\ref{RMTdef}).
\bibitem{ScalingComm} This scaling helps to single out two different parameters of
                     the problem: the small  parameter $ \, {\cal B} \ll 1 \, $ and an arbitrary ratio
                     $ \, \Omega/ {\cal B} $. To this end we can scale by the constant either
                     all variables of $ \, \Psi $-vectors, Eqs.(\ref{S-vevtors}), or  only variables 
                     $ \, \lambda_{R,A} \, $ of $ \, Q $-matrices parameterized in accordance 
                     with Appendix \ref{Q-param}. In both cases the integration measure is invariant  
                     due to the equal number of  commuting and anticommuting variables in the
                     former case and due to the formula (\ref{NewMeasure}) in the latter one.
\bibitem{Styp} This is obvious if one deforms the integration contour for the $ \, S $-variable 
                           from the real axis $ \, S \in [ 0, + \infty ] \, $ to the imaginary one $ \, S \in [ 0, 
                           + \imath \infty ] $.
\bibitem{SaddleCorr}  Small corrections to the saddle-point integration in Eq.(\ref{S-R-int}) 
                   have been obtained by a rough upper estimate. More detailed analysis shows 
                   that they can be parametrically smaller. For example, the corrections $ \, 
                   O(\sqrt{b_{p\alpha}}) \, $ cancel out in the leading term of the virial expansion 
                   for the level compressibility and only the smaller ones $ \, O( b_{p \alpha} ) \, $ 
                   remain. In such cases, validity of the saddle-point integration becomes even broader
                   than the range described by Eq.(\ref{ValidEstim}).
\bibitem{Errat} Note that the unfolding factor $ \, \sqrt{2} \, $ has been calculated in 
                            [\onlinecite{Trotter}] for the level compressibility in the case m=2 and 
                            it agrees with Eq.(\ref{Gm-Cent}). However in the subsequent paper 
                            [\onlinecite{Chi-Virial}] the same unfolding factor has been erroneously 
                            used for m=3. The correct unfolding factor is $ \, \sqrt{3} $, cf. 
                            Eq.(\ref{Gm-Cent}).
\end{thebibliography}
\end{document}